\documentclass[11pt,twoside,letterpaper]{article} %% The same for {book}
\usepackage{times,fancyhdr,graphicx}
\usepackage{url}
\def\aap{A\&A\,  }%% Astronomy and Astrophysics
\def\aj{AJ  }%% The Astronomical Journal
\def\aplett{Astrophys. Lett.\,  }%Astrophysics Letters
\def\apj{ApJ\,  }%% Astrophysical Journal
\def\apjl{ApJ\,  }%% Astrophysical Journal, Letters
\def\apjs{ApJS  }%% Astrophysical Journal, Supplement
\def\araa{ARA\&A  }% Annual Review of Astron and Astrophys
\def\mnras{MNRAS\,  }%% Monthly Notices of the RAS

\def\pasj{PASJ\,  }%% Publications of the ASJ
\def\rmp{Rev. Mod. Phys.  }% Reviews of Modern Physics
 
\def\sovast{Soviet Astronomy} %Soviet  Astronomy
  

\makeatletter
\def\cleardoublepage{\clearpage\if@twoside \ifodd\c@page\else%
    \hbox{}%
    \thispagestyle{empty}%
    \newpage%
    \if@twocolumn\hbox{}\newpage\fi\fi\fi}
\makeatother

\renewcommand{\thesection}{\arabic{section}.}
\renewcommand{\thesubsection}{\thesection\arabic{subsection}.}
\renewcommand{\thesubsubsection}{\thesubsection\arabic{subsubsection}.}

\def\figurename{Figure}
\makeatletter
\renewcommand{\fnum@figure}[1]{\figurename~\thefigure.}
\makeatother

\def\tablename{Table}
\makeatletter
\renewcommand{\fnum@table}[1]{\tablename~\thetable.}
\makeatother

\begin{document}
\title{
{\begin{flushleft} \vskip 0.45in
{\normalsize\bfseries\textit{Chapter~1}}
\end{flushleft}
\vskip 0.45in \bfseries\scshape  
The Milky Way as modeled by Percolation and Superbubbles 
}}
\author{\bfseries\itshape  
 Lorenzo Zaninetti \thanks{Email address: zaninetti@ph.unito.it}\\
Physics Department,
 via P.Giuria 1,\\ I-10125 Turin,Italy 
}

\date{}
\maketitle \thispagestyle{empty} \setcounter{page}{1}
% ------- [First Page Running Head] - place it immediately after title! ------
\thispagestyle{fancy} \fancyhead{}
\fancyhead[L]{In: Book Title \\
Editor: Editor Name, pp. {\thepage-\pageref{lastpage-01}}} % needs \label{lastpage-01} on the last page.
\fancyhead[R]{ISBN 0000000000  \\
\copyright~2007 Nova Science Publishers, Inc.} \fancyfoot{}
\renewcommand{\headrulewidth}{0pt}
%------------------------------------------------------------------------------

\vspace{2in}

\noindent \textbf{PACS} 98.35.-a ; 98.38.Mz
\vspace{.08in} \noindent \textbf{Keywords:} 
Characteristics and properties of the Milky Way galaxy;
Supernova remnants
% ------------ [Running Heads - for odd and even pages] - please insert it only on page 2!
\pagestyle{fancy} \fancyhead{} \fancyhead[EC]{Zaninetti}
\fancyhead[EL,OR]{\thepage} \fancyhead[OC]
{
Percolation and Superbubbles
} \fancyfoot{}
\renewcommand\headrulewidth{0.5pt}

\section*{Abstract}

The spiral structure of the Milky Way 
can be  simulated by adopting 
percolation theory, where  the active zones are
produced  by  the evolution of many supernova (SN).
Here we assume conversely  
that the  percolative process 
is triggered by superbubbles (SB), the  result 
of  multiple SNs.
A first thermal model takes into account
a bursting phase  which evolves in a medium with 
constant density, and a subsequent 
adiabatic expansion which  evolves in a medium 
with decreasing density along the galactic height.
A second cold model follows the evolution 
of an SB in an auto-gravitating medium 
in the framework of the momentum conservation 
in a thin layer.
Both the thermal and cold models are compared 
with the results of numerical 
hydro-dynamics.
A simulation  of GW~46.4+5.5, 
the  Gould Belt,
and    
the   Galactic Plane is reported. 
An elementary theory of the image, which  allows 
reproducing the hole visible at the center 
of the observed SB, is provided.

\section{Introduction}

The term  supershell  was observationally defined 
by \cite{heiles1979},
where eleven H\,I objects were examined.
Supershells   have been observed  as   expanding  shells,
or holes,  in the H\,I-column density   distribution  of  our galaxy,
in the   Magellanic Clouds,
in   the dwarf irregular
galaxy Ho-II \cite{Puche1992},
and in the nearby  dwarf galaxy IC 2574 \cite{Walter1998}.
The dimensions of these objects  span from 100~\mbox{pc}
to 1700~\mbox{pc}
and often present  elliptical shapes or elongated features, which
  are difficult to explain based on   an expansion in a uniform
medium.
These structures  are commonly  explained through introducing
theoretical objects named  bubbles or  superbubbles (SB);
these  are created by  mechanical energy input from stars 
(see  for example \cite{PikelNer1968}; \cite{weaver}).
Thus, the origin of a supershell is not necessarily an
   SB, and other
   possible origins include collisions of high-velocity clouds (see
   for example \cite{Tenorio-Tagle1988}; \cite{Santillian1999}).
   A  worm is another 
    observed feature that may, or may not, be associated with a wall
    of an SB.
    Galactic worms were 
    first identified as irregular, vertical columns of
    atomic gas stretching from the galactic plane;
    now, similar structures are found
    in radio continuum and infrared maps 
     (see for example \cite{Koo1992}). 
SBs are also observed in other galaxies,
we cite two examples among many:
a kinematical and photometric catalog 
was compiled for 210 H II regions in the Large Magellanic
Cloud \cite{Ambrocio-Cruz2016} and
12 SBs were classified in the irregular galaxy  NGC 1569 
\cite{Sanchez-Cruces2015}.

The models 
that  explain  SBs  as being due to 
the combined  explosions  of supernova in a cluster  of
massive stars are now briefly reviewed.
Semi-analytical and hydro-dynamical
calculations are generally adopted.
In  semi-analytical calculations,
the thin-shell approximation
can be the key to obtaining
the expansion of
the SB; see,  for example,
\cite{McCray1987,mccrayapj87,MacLow1988,Igumenshchev1990,Basu1999}.

   The thin-shell approximation has already been  used in a variety of
   different problems, and with both analytical and numerical approaches
   (besides the review by 
         \cite{Bisnovatyi-Kogan1995}, see 
         \cite{Begelman1992} and  
         \cite{Moreno1999}). Thus 
   the validity and limitations of the method are well known. For instance,
   modeling is fast and simple because the shell dynamics are only included
   in an approximate way, while the fluid variables are not included
   at all. 
  The price
  that one has to pay is a lack of knowledge of  the density and
  velocity profiles and, obviously,  neither the onset and
  development
  of turbulence, instabilities, or mixing  can be followed. 
  These facts limit the applications
  of this method to deriving only the general shapes, approximate expansion
  rates, and gross features of the surface mass-density distributions 
  of the
  shells.

The hydro-dynamical approximation was used by 
\cite{MacLow1989}.
As for the effect of magnetic fields,
a  semi-analytical method was  introduced
by \cite{Ferriere1991} and
a  magneto-hydrodynamic 
code has been adopted  by various authors
\cite{Tomisaka1992,Tomisaka1998,Kamaya1998}.
The often adopted exponential and vertical profiles in density  
do not correspond
to a physical process of equilibrium.
The case of an isothermal self-gravitating disk (ISD) 
is  an equilibrium  vertical profile which can be 
coupled with momentum
conservation.
The models cited leave some questions
unanswered or only partially answered:
\begin {itemize}
\item  Is it possible to calibrate the vertical profile 
       of an ISD?
\item  Is it possible to deduce an analytical formula for 
        the temporal evolution of an SB in the presence 
        of a vertical profile density as given by an ISD?
\item Is it possible to  deduce numerical results for an SB when 
      the  expansion starts at a given galactic height?
\item What is  the influence of galactic rotation 
      on the temporal evolution of an SB?
\item Can we explain the worms as a particular effect using 
      image   theory   applied to SBs?
\end{itemize}

In order  to answer these questions,
Section \ref{secpercolation} 
reviews the percolation theory  
which allows modeling a spiral galaxy.
Section \ref{sectionthermal} 
introduces a  numerical code which solves the 
momentum equation coupled
with  the variation of pressure in the presence
of the injection of
mechanical luminosity and the adiabatic losses,
the so called {\it thermal model}.
Section \ref{sectioncold} 
models an aspherical expansion,
in 
a vertical  profile   in the number of particles
as  given by an ISD,
the so called {\it cold model}.
Section \ref{sectionhydro} 
reports a comparison between the 
numerical hydro-dynamical 
calculations and the two codes here presented.
Section \ref{sectionastro}
models  the SBs associated with  GW~46.4+5.5, 
the  Gould Belt,
and    
the   Galactic Plane. 
Section \ref{sectionimage}
contains detailed information
on how to build an image of an SB 
both in the symmetric and asymmetric case. 

\section{The Percolation}
\label{secpercolation}
The appearance of  arms can be simulated through
percolation
theory 
 \cite{Seiden1979,Seiden1983,Schulman1986,Zaninetti1988,Seiden1990}.
The fundamental   hypotheses
and   the parameters adopted in the
simulation will now be reviewed.
\begin{enumerate}
\item
The motion of a gas on the galactic plane has a constant
rotational  velocity, denoted by
${\it V}_{\mathrm{G}}$ (in the case of
spiral type Sb     218 $\mathrm {km\,s^{-1}}$).
Here the velocity, ${\it V}_{\mathrm {G}}$, is
expressed in units of 200 $\mathrm {km\,s^{-1}}$ and will therefore be
${\it V_{\mathrm{G}}}$ =1.09.
\item
The polar simulation array   made by rings and cells
has  a radius
${\it R}_{\mathrm{G}}$ = 12~\mbox{kpc}.
The  number of rings, (59),
by   the multiplication of  ${\it R}_{\mathrm{G}}$
with  the number of rings for each
kpc, denoted by  \mbox{nring/kpc},
which in our case is  5.
Every ring is then made up of
many~{\it cells}, each one with a  size on  the
order of the galactic thickness,
 $ \approx$ 0.2 \mbox{kpc}.
The parameter  \mbox{nring/kpc} can  also be found
by
dividing {\mbox  1 kpc} by the cell's approximate size.

\item  The global number of cells, 11121,
multiplied by the probability
of spontaneous new cluster formation,
$p_{\mathrm {sp}}$ (for example 0.01),
 allows the process to start (with the previous parameters,
 111 new clusters were generated).
Each one of these
sources has six new surroundings that are labeled for each ring.
\item   In order to better simulate the decrease of the gas density
along the radius,  a stimulated  probability of forming
new clusters
with  a linear dependence by  the radius
,$ p_{\mathrm st}$ = $ a + b R $, was chosen.
The values $a$ and $b$ are found by fixing ${\it prmax}$
( for example 0.18),
 the stimulated probability
at the outer ring, and ${\it prmin}$  (for example 0.24)
 in the inner ring;
of course,
{\it prmin}$\geq ${\it prmax}.
This  approach  is surprisingly similar to the introduction
of  an anisotropic
probability distribution in order to better
simulate certain classes of spirals
\cite{Jungwiert1994}.
\item   Now, new sources are selected in
each ring based on  the hypothesis of  different
stimulated probabilities.
A rotation curve is imposed
so that the array rotates in the same  manner
as the  galaxy. 
The
procedure repeats
itself $n$ times (100); we denote by ${\it t}_{\mathrm{G}}$ the
age  of the simulation, viz., $100\cdot10^7 \mbox {yr}$,
where  $10^7$ \mbox{yr} is  the
astrophysical counterpart of one time  step.
\item In order to prevent catastrophic growth, the process
is stopped when the number of surroundings is
greater than ${\it max}$ (1000)
and restarts by spontaneous probability.
\item The final number of active cells  (3824)
is plotted with the size, 
which  decreases linearly  with the cluster age.   In
other words,  
the young clusters are bigger than the old ones.
Only ten
cluster ages  are shown;
only  cells with an
age of less than ${\mathrm{ life}}$  (in our case $10\cdot10^7$ yr)
are selected.
A more deterministic  approach has been followed 
by \cite{Palous1994},
where  a model that connects 
the energy injection by star  formation with the resulting
interstellar  structures in a differentially rotating disc  has been
introduced. 
\end{enumerate}

A typical run is shown in Figure \ref{secpercolation} and in 
Table \ref{tablemilky} we summarize 
the parameters used in this simulation.
%begin figure milky
\begin{figure}
\begin{center}
\includegraphics[width=7cm]{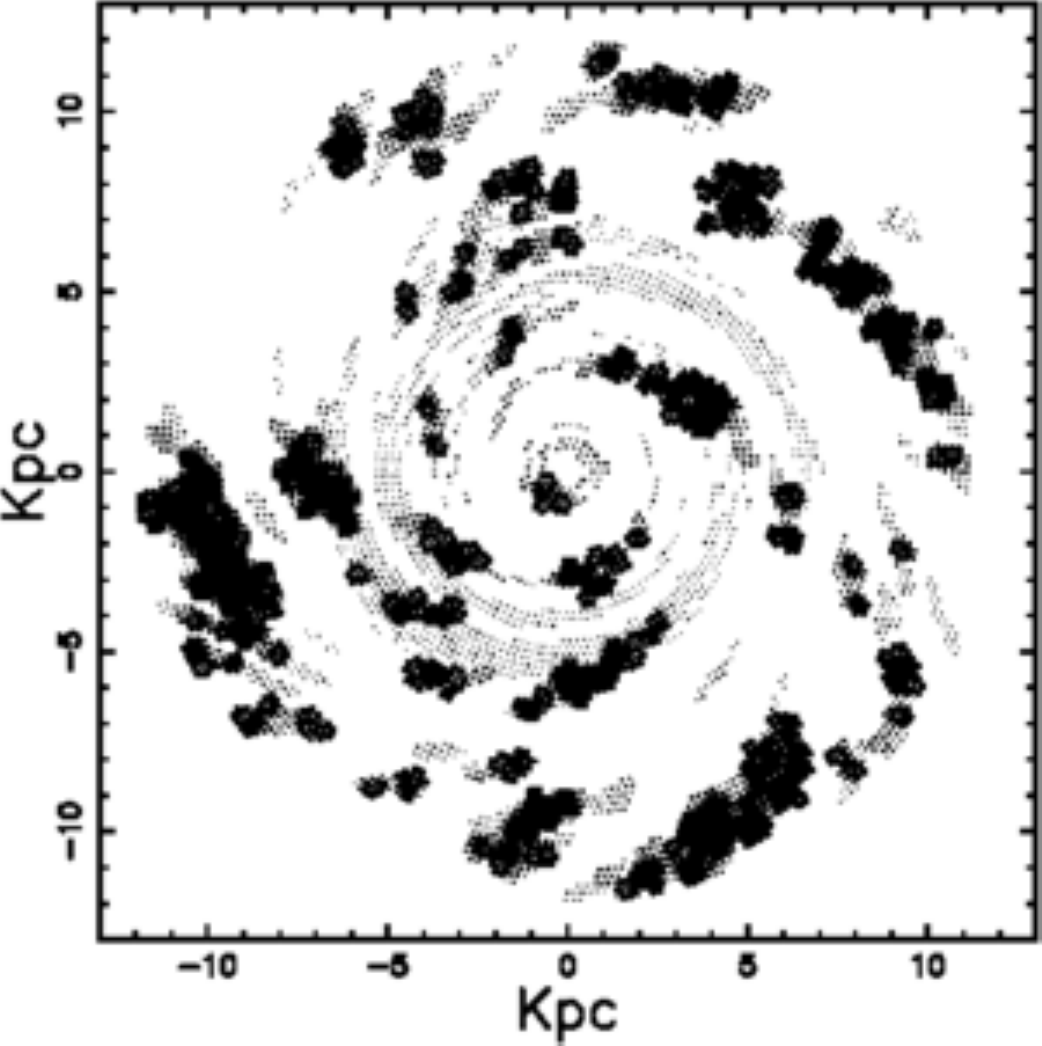}
\end{center}
\caption{
A typical simulation of the galaxy's arms. 
Here `active' is the final
number of active cells, `random' the initial number of randomly
distributed sources, `pc' the considered probability,
`cells' the total number of cells,
  $R$ the radius in Kpc, `ring/kpc' the number
of rings considered per kpc,  the size of the cells is in kpc,
$v$ the velocity in units of 200 Km/sec, and the time is expressed 
in units of millions of years, see the numerical values 
in Table \ref{tablemilky}
}
 \label{milky}%
\end{figure}
% end figure milky

\begin{figure}
\begin{center}
\includegraphics[width=7cm]{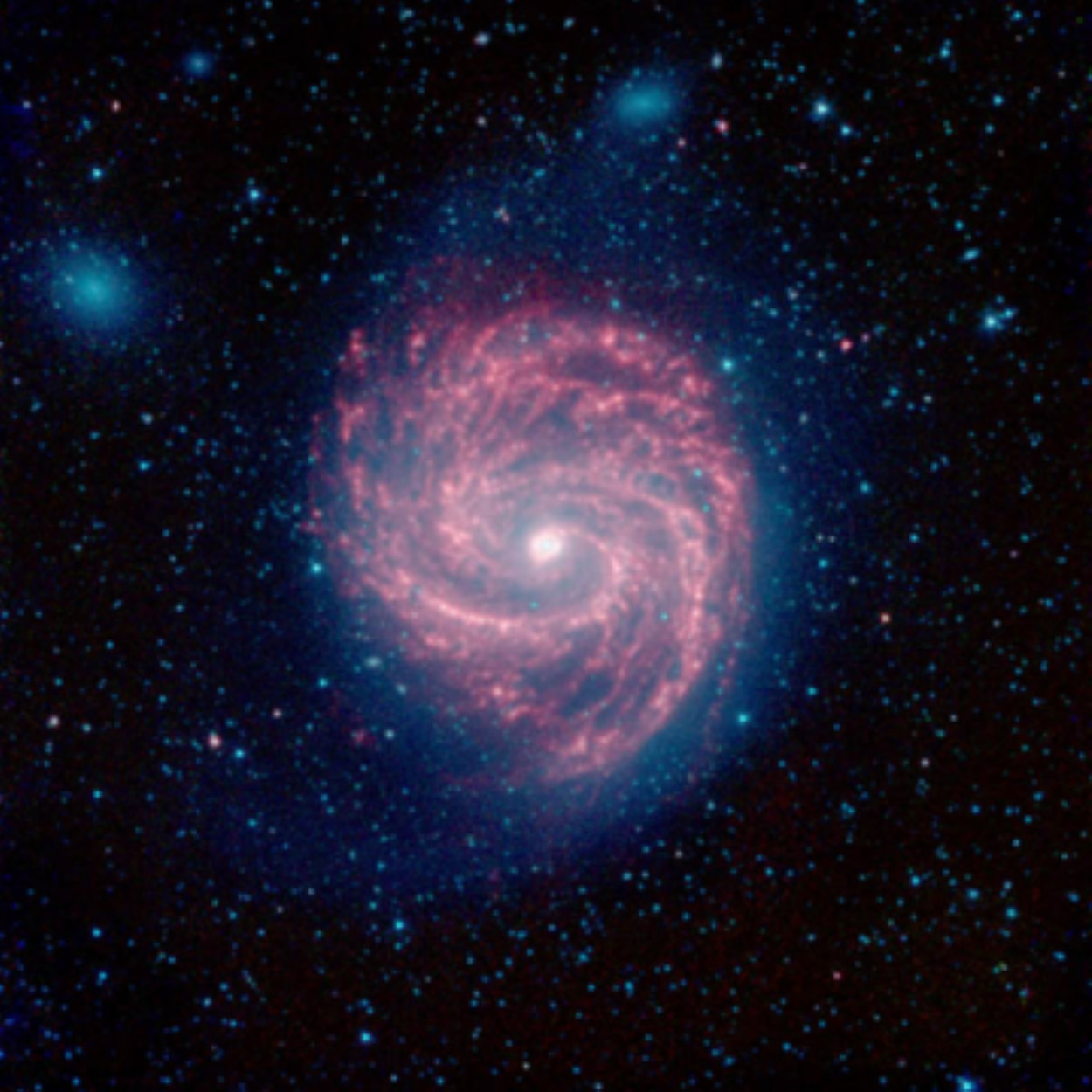}
\end{center}
\caption
{
Composite image in the infrared of M100.
}
 \label{m100}%
\end{figure}

\begin{table}
\caption
{
Parameters of the model
}
 \label{tablemilky}
 \[
 \begin{array}{cc}
 \hline
 \hline
 \noalign{\smallskip}
 galactic~ radius              & 12 kpc              \\
 rings/kpc                     & 5                   \\
 stochastic~ probability        & 0.01                \\
 timestep                      & 10^7\, years        \\
 cells                         & 11121               \\
 active                        & 3894                \\
 \hline
 \hline
 \end{array}
 \]
\end{table}
 A real spiral galaxy as  observed in the infrared is reported 
in Figure \ref{m100} as observed by the Spitzer Space Telescope,
more details are available at  \url{http://www.spitzer.caltech.edu/}.

%begin figure m100
% end figure m100

\section{The thermal model for SBs}
\label{sectionthermal}
The starting equation for the evolution
of the SB \cite{McCray1987,mccrayapj87,Zaninetti2004}
 is  momentum conservation
 applied to a pyramidal section,   characterized
by a solid angle $\Delta \Omega_j$:
\begin {equation}
\frac{d}{dt}\left(\Delta M_j \dot{R}_j\right)=p R_j^2
\Delta \Omega_j
\quad ,
\label{eq_momentum}
\end {equation}
where the pressure of the surrounding medium is
assumed to be negligible and the mass
is confined in a thin shell with mass $\Delta M_j$.
The subscript $j$  was added  here
in order to note
that this is not a spherically symmetric system.
Due to the fact that  $p$ is uniform in the cavity of
the SB, a summation to obtain the
total volume, such as
\begin{equation}
V=\sum  \Delta \Omega_j R_j^3/3
\label{eq_volume}
\quad ,
\end  {equation}
is necessary  to determine its value.
The mass conservation equation for a thin
shell is
\begin {equation}
\Delta M_j = \frac{1}{3}R_j^3 \bar{\rho}_j\Delta\Omega_j
 \quad .
\end {equation}
The  pressure is enclosed in the energy conservation equation,
\begin{equation}
\frac{1}{\gamma-1}\frac{d(pV)}{dt}=L-p\frac {dV}{dt}
\quad 
,
\label{eqn:MMenergy}
\end{equation}
where $L=E_0R_{\rm SN}/4\pi$ denotes
the mechanical luminosity
injected into  a
unit
solid angle and $\gamma~=5/3$.
Equation~(\ref{eqn:MMenergy}) can be expanded,
obtaining
\begin {equation}
\frac {dp}{dt} = \frac { L (\gamma -1 )} {V}
            - \gamma  \frac {p}{V} \frac {dV}{dt}
\label{eqn:dpdt}
\quad  .
\end {equation}
Formulae~(\ref{eq_momentum}) and  (\ref{eqn:dpdt})
will be our basic equations to  numerically integrate 
in the bursting phase.
An approximation concerning the pressure can  be obtained by
ignoring the second  term of the rhs in~(\ref{eqn:MMenergy});
this leads to
\begin {equation}
p=\frac{2E_0R_{\mathrm{SN}}t}{3V}
\quad ,
\label {eq:pressure}
\end {equation}
where  $t$ is the considered time, $R_{\mathrm{SN}}$
the rate of supernova explosions, $E_0$ the energy of each
supernova, and $V$   is computed as in  
Equation~(\ref{eq_volume}).
On   continuing  to   consider   the case of constant density,
the volume becomes
\begin {equation}
V=\frac{4\pi}{3}R^3
\quad 
 .
\end {equation}
Equations (\ref{eq_momentum}) and (\ref{eq:pressure})
 lead to
\begin {equation}
\frac{d}{dt}\left(\bar{\rho}R^3\dot{R}\right)=
\frac{3E_0R_{\mathrm{SN}}}{2\pi}
\frac{t}{R}
 \quad .
\label {eq:firstder}
\end {equation}
In order to integrate the above equation,
$R^3 \dot{R}=AR^\alpha$
is imposed.
After adopting the initial condition of  $R=0$ at $t=0$
and assuming
$\bar{\rho}$ is constant irrespective of $R$ or $t$,
the equation
for the expansion speed is obtained:
\begin {equation}
\dot{R}=\frac{\alpha+1}{\alpha}
 \frac{3E_0R_{\mathrm{SN}}}{4\pi \bar{\rho}R^4}t^2
\quad  .
\label{approximation}
\end {equation}
By integrating  the previous equation,
 ``the expansion law'' is obtained:
\begin {equation}
R=\left[\frac{5(\alpha+1)}{4\pi \alpha}\right]^{1/5}
  \left(\frac{E_0R_{\mathrm{SN}}}{\bar{\rho}}\right)^{1/5} t^{3/5},
\label{eq:firstradius}
\end {equation}
which is identical to equation (10.34) of
 \cite{McCray1987},
 since $\alpha=7/3$.

\subsection{After SN Bursts Stop}

\label{approximation2}
It is clear that  an upper limit
should be inserted into
the basic equation~(\ref{eq:pressure}); this  is the time
after which the bursting phenomenon stops, 
${t^{\mathrm{burst}}}$.

Since there is a $pdV$ term in the the First Law
of Thermodynamics ($dQ=0=dU+pdV$),
the total thermal energy decreases with time.
The pressure of  the internal gas decreases according to
the adiabatic law,
\begin {equation}
p=p_1({V_1 \over V})^{5/3}
\label {eq:adiabatic1}
\quad
,
\end {equation}
where
\begin {equation}
p_1 =  \frac {2}{3}  \frac {E_0R_{\mathrm{SN}} t^{\mathrm{burst}}}
    {  V_1}
\quad 
,
\label {eq:adiabatic3}
\end {equation}
and the equation for
the conservation of  momentum
becomes
\begin {equation}
\frac{d}{dt}\left(\bar{\rho}_j~R_j^3\dot{R_j}\right)=
3 \frac{p_1 V_1^{5/3} }{ V^{5/3}   }
R_j^2
\quad 
 .
\label {eq:adianum}
\end {equation}
It is important to remember that this phase 
occurs after the SN burst stops.
At that time ($t=t^{\mathrm{burst}}$), 
the volume  of the bubble
is computed  by using
$R_j(t^{\mathrm{burst}})$,
 and the expansion speed is equal
to $\dot{R}_j(t^{\mathrm{burst}})$,
both of which are obtained from
the numerical solution of
Equation (\ref{eq:firstder}).
On assuming  (also here)   that
$\bar{\rho}$ is constant irrespective of $R$ or $t$,
Equation~(\ref{eq:adiabatic3}) is now inserted into
Equation~(\ref {eq_momentum}) and
  the expansion law  is obtained for the after-burst phase,
\begin {equation}
R=\left(\frac{147} {4\pi }\right)^{1/7}
  \left(\frac{E_0R_{\mathrm{SN}} t^{\mathrm{burst}} R_1^2 }
    {\bar{\rho}}\right)^{1/7} t ^{2/7}
\quad  
 ,
\label{eq:secondstradius}
\end {equation}
which is identical to  equation (10.33) of
 \cite{McCray1987}.

\subsection{Numerical Integration}

\label{sectheta}
The differential  equations need to be solved by sectors,
with each sector being treated  as independent from the others,
except for the coupling in computing 
the  volume of the bubble,  
see Equation (\ref{eq_volume}).
The
range   of the polar angle $\theta$ ($180^{\circ}$)
will be divided
into  $n_{\theta}$ steps, and the range of the azimuthal angle
$\phi$ ($360^{\circ}$) into  $n_{\phi}$   steps.
This will  yield  ($n_{\theta}$ +1) ($n_{\phi}$ +1) directions of
motion that can  also  be identified with  the number of vertices
of the polyhedron representing the volume occupied by the
expansion;
this   polyhedron  varies  from  a sphere
to various morphological  shapes  based on  the
swept up  material
in each direction.
In  3D plots showing the  expansion surface  of the explosion,
the number  of vertices is    $n_{\mathrm {v}}$ =
\label{sec_nv}
($n_{\theta}$+1)$\cdot$($n_{\phi}$ +1)
and  the number
of  faces is 
$n_{\theta}\cdot\;n_{\phi}$,
typically we have     $n_{\theta}$=50
and $n_{\phi}$=50 (in this case the subscript  varies
between 1 and 2601).
However, all calculated  models  are axisymmetric,
and the essential number of points to draw such
figures is only $n_\theta+1$=51.
 $R_{\mathrm{up}}$,
$R_{\mathrm{eq}}$ and  $R_{\mathrm{down}}$  are now introduced, which
represent  the distances  from the  position of  the
the OB associations  (denoted by $z_{\mathrm{OB}}$)
to the top, to the left and  to
the bottom of the bubble.

\subsection{The Numerical Equations}

At each  time step, $\Delta t$,   the volume {\it V}  of the
expanding bubble is computed, see Equation (\ref{eq_volume}).
In  other words,    the volume {\it  V }
swept up   from the explosions  is no  longer a sphere,
but  becomes an egg or an hourglass.
The pressure in the first phase,
see Equation (\ref{eqn:dpdt}),  is computed through the
following finite-difference approximation:
\begin{equation}
p^k  =
p^{k-1} + \left [  L \frac {\gamma -1} {V^k}
- \gamma \frac {p^{k-1}}{V^k}  \frac{V^k -V^{k-1}} {\Delta t}
\right ]  \Delta t
\quad 
,
\end{equation}
where $k$ is the number of steps employed.

Equation~(\ref{eq_momentum})
now leads  to
\begin {equation}
\frac{d}{dt}\left(\bar{\rho_j} R_j^3\dot{R_j}\right)=
3 pR^2_j
\quad 
 ,
\label {eq:firsnum}
\end {equation}
which  may  rewritten as
\begin {equation}
3 R_j^2 \dot{R_j}^2 + R_j^3 \ddot {R_j} =
3p \frac{R^2_j}{\bar{\rho_j}}
- \frac{\dot{ \bar{\rho_j}} }{\bar{\rho_j}} R_j^3\dot{R_j}
\quad  .
\label {eq:secondnum}
\end {equation}
The first term represents the ram pressure of the stratified ISM
on the expanding surface and the second represents the inertia
of the bubble. The average density $\bar{\rho_j}$ is
numerically computed according to the algorithm outlined
in subsection \ref {sec_rhoaverage}; the time derivative of
$\bar{\rho_j}$ at each time step $\Delta t$
 is computed according
to the finite difference-approximation,
\begin{equation}
\dot{ \bar{\rho_j}} =
\frac { \bar{\rho_j}^k -\bar{\rho_j}^{k-1}}
{\Delta t}
\quad ,
\label{deriv_rho}
\end{equation}
where {\it k } is the number of steps considered.

Equation~(\ref{eq:secondnum})  can be re-expressed
as two  differential
equations (along each direction $j$)
of the first order suitable to be integrated:
\begin {equation}
 \frac {dy_{1,j}} {dt}   = y_{2,j}
,
\end  {equation}
\begin{equation}
\frac  {dy_{2,j}}{dt}    =
3p \frac{ 1  }{ \bar{\rho_j} y_{1,j}}
-3 \frac { y_{2,j}^2 }{ y_{1,j}}
- \frac{\dot{ \bar{\rho_j}} }{\bar{\rho_j}} y_{2,j}
\quad  
.
\label {eq:twonum}
\end {equation}

In the new  regime ($t \geq t^{\mathrm{burst}}$),
Equation~(\ref{eq:adianum}) now becomes
\begin {equation}
3 R_j^2 \dot{R_j}^2 + R_j^3 \ddot {R_j} =
3 \frac{p_1 V_1^{5/3} }{ V^{5/3}   \bar{\rho_j}}
R_j^2
- \frac{\dot{ \bar{\rho_j}} }{\bar{\rho_j}} R_j^3\dot{R_j}
 \quad .
\label {eq:numericaladia}
\end {equation}
Equation~(\ref{eq:numericaladia})  can be re-expressed
as  two  differential
equations (along each direction {\it j})
of the first order suitable to be integrated:
\begin {equation}
 \frac {dy_{1,j}}{dt}   = y_{2,j}
\quad 
,
\end  {equation}
\begin{equation}
\frac  {dy_{2,j}}{dt}   =
\frac{3 p_1 V_1^{5/3}} { V^{5/3} \bar{\rho_j} y_{1,j}}
-3 \frac { y_{2,j}^2 }{ y_{1,j}}
- \frac{\dot{ \bar{\rho_j}} }{\bar{\rho_j}} y_{2,j}
\quad  
.
\label {eq:twonumadia}
\end {equation}

The  integrating scheme  used  is the Runge--Kutta  method,
and   in particular the  subroutine RK4
(\cite{press}); the  time  derivative of the  density
along a certain direction $j$, $\dot{ \bar{\rho_j}}$,
is computed at each time step
according to formula~(\ref{deriv_rho}).

The integration time, $t_{\mathrm{age}}$,  and the time steps,
are always  indicated in the  captions
of the various diagrams.
The pressure
across the bursting time  
is continuous. 
This  is because the first value  of the pressure
after the bursting time  is that  of the last step
in the bursting phase,  modified according to the  
adiabatic law   modelled  by  formula~(\ref{eq:adiabatic1}).

The upper limit  chosen to integrate the  differential 
equations is  $t_{\mathrm {age}}=2.5\cdot10^7~{\mathrm{yr}}$;
this is the approximate age of GSH~238.

\subsection{The Density Profile}

\begin{figure}
  \begin{center}
\includegraphics[width=7cm]{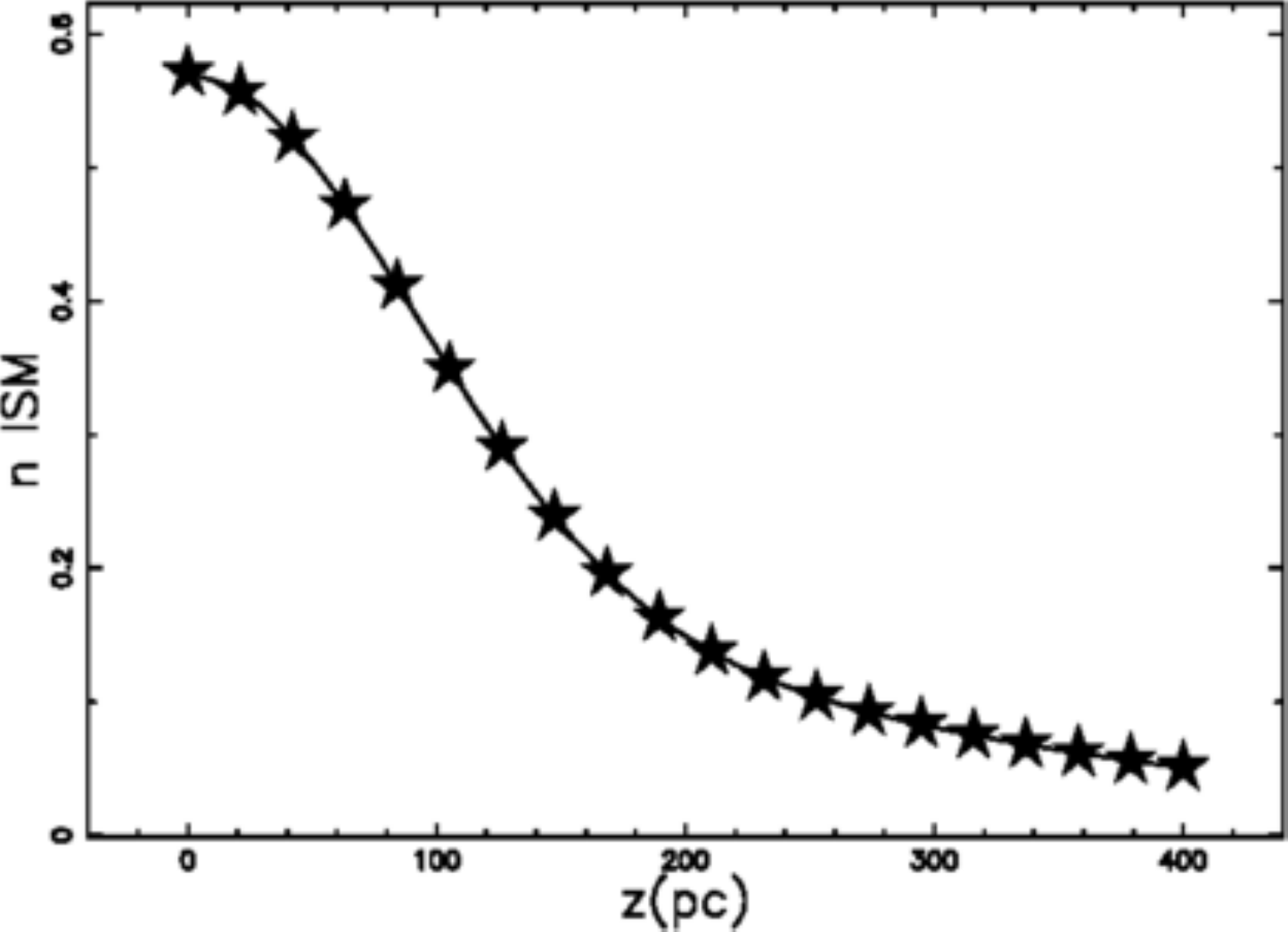}
  \end{center}
\caption{
The average structure of the gaseous disk in the {\it z}-direction;
{\it z}  is allowed to vary between 0 \mbox{pc} and  400 \mbox{pc}.
}
\label{figism}%
    \end{figure}

The vertical density distribution
of galactic H\,I  is well-known; specifically, it has the following
three component  behavior as a function of
{\it z}, the  distance  from  the galactic plane in pc:
\begin{equation}
n(z)  =
n_1 e^{- z^2 /{H_1}^2}+
n_2 e^{- z^2 /{H_2}^2}+
n_3 e^{-  | z |  /{H_3}}
\,.
\label{equation:ism}
\end{equation}
\label{Sec_ISM}
We took \cite{Bisnovatyi1995,Dickey1990,Lockman1984} 
 $n_1$=0.395 ${\mathrm{particles~}}{\mathrm{cm}^{-3}}$, $H_1$=127
        \mbox{pc},
        $n_2$=0.107 $\mathrm{particles~}{\mathrm{cm}^{-3}}$, $H_2$=318
        \mbox{pc},
        $n_3$=0.064 $\mathrm{particles~}{\mathrm{cm}^{-3}}$, and  $H_3$=403
        \mbox{pc}.
This  distribution  of  galactic H\,I is valid in the range
0.4 $\leq$  $R$ $\leq$ $R_0$, where  $R_0$ = 8.5 \mbox{kpc}
and $R$  is the
distance  from  the center of the galaxy.
A plot  showing such a dependence of  the ISM 
density 
on
{\it z}  is shown in Figure \ref{figism}.

\subsection{Computation of the Swept Mass}

The  ISM density is not constant, but  varies  in the
{\it z}-direction according  to  Equation~(\ref{equation:ism}).
The swept mass can therefore be computed in a
certain direction {\it j}  using the
following  algorithm:
\begin{enumerate}
\item
The pyramidal sector is divided into layers (for example 1000)
whose radii range from $R_{L-1/2}$ to $R_{L+1/2}$.
\item  In each layer  the volume
$
\Delta V_L=\frac{1}{3}(R_{L+1/2}^3-R_{L-1/2}^3)
$
is computed as well  as the corresponding   mass,
$\Delta M_L=\Delta V_L\rho_j(z=R_L\cos \theta_j+z_{\mathrm{OB}})$,
where $\theta_j$ represents an angle between the $z$-axis and the $j$th
radial path.

\item The various contributions, $\Delta M_L$,
are added in order to obtain
the total swept  mass in the considered sector.

\item
The average density along a sector {\it j}
 as $\bar{\rho_j}$ is calculated
 using $\Delta M_L$ and $\Delta V_L$ as $\bar{\rho_j}=\sum_L \Delta M_L/
\sum_L\Delta V_L$.

\end{enumerate}

\label {sec_rhoaverage}

\subsection{Astrophysical Units}

\label{astrounits}
Our basic units are: time ($t_7$), which
is expressed  in units of $10^7$ \mbox{yr};
$E_{51}$, the  energy in  $10^{51}$ \mbox{erg};   $n_0$, the
density expressed  in particles~$\mathrm{cm}^{-3}$~
(density~$\rho_0=n_0$m, where m=1.4$m_{\mathrm {H}}$); and
$N^*$,  which  is the number of SN explosions
in  $5.0 \cdot 10^7$ \mbox{yr}.

By using the previously defined  units,
formula~(\ref {eq:firstradius}) concerning
``the expansion law'' in the bursting phase  becomes
\begin{equation}
R =111.56\;  \mathrm{pc}(\frac{E_{51}t_7^3 N^*}{n_0})^{\frac {1} {5}}
.
\label{eqn:raburst}
\end{equation}

Conversely, Equation~(\ref {eq:secondstradius}) concerning
  ``the expansion law'' in the adiabatic phase
is
\begin{equation}
R =171.85\;  \mathrm{pc}( \frac {E_{51}N^*} {n_0} )^{\frac {1}{5}}
                   (t^{\mathrm{burst}}_7)^{\frac {11}{35}}
                   (t_7)^{\frac {2} {7} }
 .
\label{eqn:raadia}
\end{equation}
This is the approximate  radius derived from a spherical model.
A perfect coincidence is not expected
upon  making  a comparison  with
that expected in
a stratified medium obtained with the multidimensional
thin shell
approximation.
Another useful formula is  the luminosity of the bubble,
see \cite{McCray1987},
\begin{equation}
L_{\mathrm{SN}}=E_0R_{\mathrm{SN}}=0.645^{36} \mbox {erg}~  \mbox{s}^{-1}E_{51}N^*
 .
\end{equation}
This formula can be useful in order to derive
the parameter  $N^*$ from observations;
 for example,
a luminosity of  $1.6~10^{38} \mbox {erg~s}^{-1}$
corresponds to  $N^*=248$.  
The total deposited energy, $E_{\mbox{tot}}$, is
\begin{equation}
E_{\mbox{tot}} = E_{51}N^* 10^{51} 
\frac {t_{\mathrm {age}}}{5 \cdot 10^7 {\mathrm{yr}}} {\mbox{erg}}
,
\label{etotage}
\end{equation}
when $t^{\mathrm{burst}}$=$t_{\mathrm {age}}$ and
\begin{equation}
E_{\mbox{tot}} = E_{51}N^* 10^{51} 
\frac {t^{\mathrm{burst}}}{5 \cdot 10^7 {\mathrm{yr}}} \mbox{erg}
,
\label{etotburst}
\end{equation}
when $t^{\mathrm{burst}}~<~t_{\mathrm {age}}$.

The spectrum of  the radiation emitted depends on
 the temperature
behind the shock front,  see  for example formula 9.14  
in \cite{mckee},
\begin{equation} 
T = \frac{3}{16} \frac{\mu} {k} v_{\mbox{s}}^2  ~~~{\mbox{K}}^{\circ}
,
\end{equation}
where $\mu$ is the mean mass per particle, {\it k}
 the Boltzmann constant
and $v_{\mbox{s}}$ the shock velocity expressed in $\mbox{cm~sec}^{-1}$.
A formula which is useful for the implementation in the code is easily
derived, 
\begin{equation}
T = 31.80 ~ v_{\mbox{sk}}^2 ~~~{\mbox{K}}^{\circ}
,
\label{formulat}
\end{equation}
when $v_{\mbox{sk}}$ is  expressed in $\mbox{ km~sec}^{-1}$. 

\section{The Cold model for SBs}
\label{sectioncold}
This Section reviews the standard thin layer 
approximation and then introduces two recursive
equations for the dynamical evolution  of a 
SB in an auto-gravitating medium, 
see \cite{Zaninetti2012g}.

\subsection{The symmetrical thin layer approximation}

\label{sec_motion}

The thin layer approximation assumes that all the swept-up 
gas accumulates infinitely in a thin shell just after
the shock front.
The conservation of radial momentum requires that 
\begin{equation}
\frac{4}{3} \pi R^3 \rho \dot {R} = M_0
\quad,
\end{equation}
where $R$ and $\dot{R}$   are  the radius and the velocity
of the advancing shock,
$\rho$ is the density of the ambient medium,
$M_0$  is the momentum evaluated at $t=t_0$,
$R_0$  is the initial radius,  
and 
$\dot {R_0}$  the  initial velocity,
see \cite{Dyson1997,Padmanabhan_II_2001}.
The law of motion is 
\begin{equation}
R = R_0 \left  ( 1 +4 \frac{\dot {R_0}} {R_0}(t-t_0) \right )^{\frac{1}{4}}  
\label{radiusm}
\quad ,
\end{equation}  
and the velocity 
\begin{equation}
\dot {R} = \dot {R_0} \left ( 1 +4 \frac{\dot {R_0}} {R_0}(t-t_0)\right )^{-\frac{3}{4}}  
\label{velocitym} 
\quad . 
\end{equation}   

\subsection{Asymmetrical  law  of motion }

\label{sec_asymmetry}

Given the Cartesian   coordinate system
$(x,y,z)$,
the plane $z=0$ will be called the equatorial plane,
$z= R \sin ( \theta) $,
where $\theta$ is the latitude angle
which has range  
$[-90 ^{\circ}  \leftrightarrow  +90 ^{\circ} ]$,
and   $R$ is the distance from the origin.
The latitude angle is  often used  in
astrophysics  to model asymmetries in
the  polar lobes,
see the example of the nebula around
 $\eta$-Carinae (Homunculus)  shown in Table 1 in  
 \cite{Smith2002}.
In our  framework,  the polar angle of the spherical
coordinate system
is  $90 - \theta$.
The vertical number density distribution
of galactic H\,I  is usually modeled by  the 
three component 
function as given by eqn. (\ref{equation:ism}).
Here, conversely,  we adopt 
the density profile of a thin
self-gravitating disk of gas which is characterized by a
Maxwellian distribution in velocity and  distribution which varies
only in the $z$-direction (ISD).
The  number density
distribution is  
\begin{equation}
n(z) = n_0 sech^2 (\frac{z}{2\,h})
\quad ,
\label{sech2}
\end{equation}
where $n_0$ is the density at $z=0$,
$h$ is a scaling parameter, 
and  $sech$ is the hyperbolic secant  
(\cite{Spitzer1942,Rohlfs1977,Bertin2000,Padmanabhan_III_2002}).

\begin{figure}
  \begin{center}
\includegraphics[width=7cm]{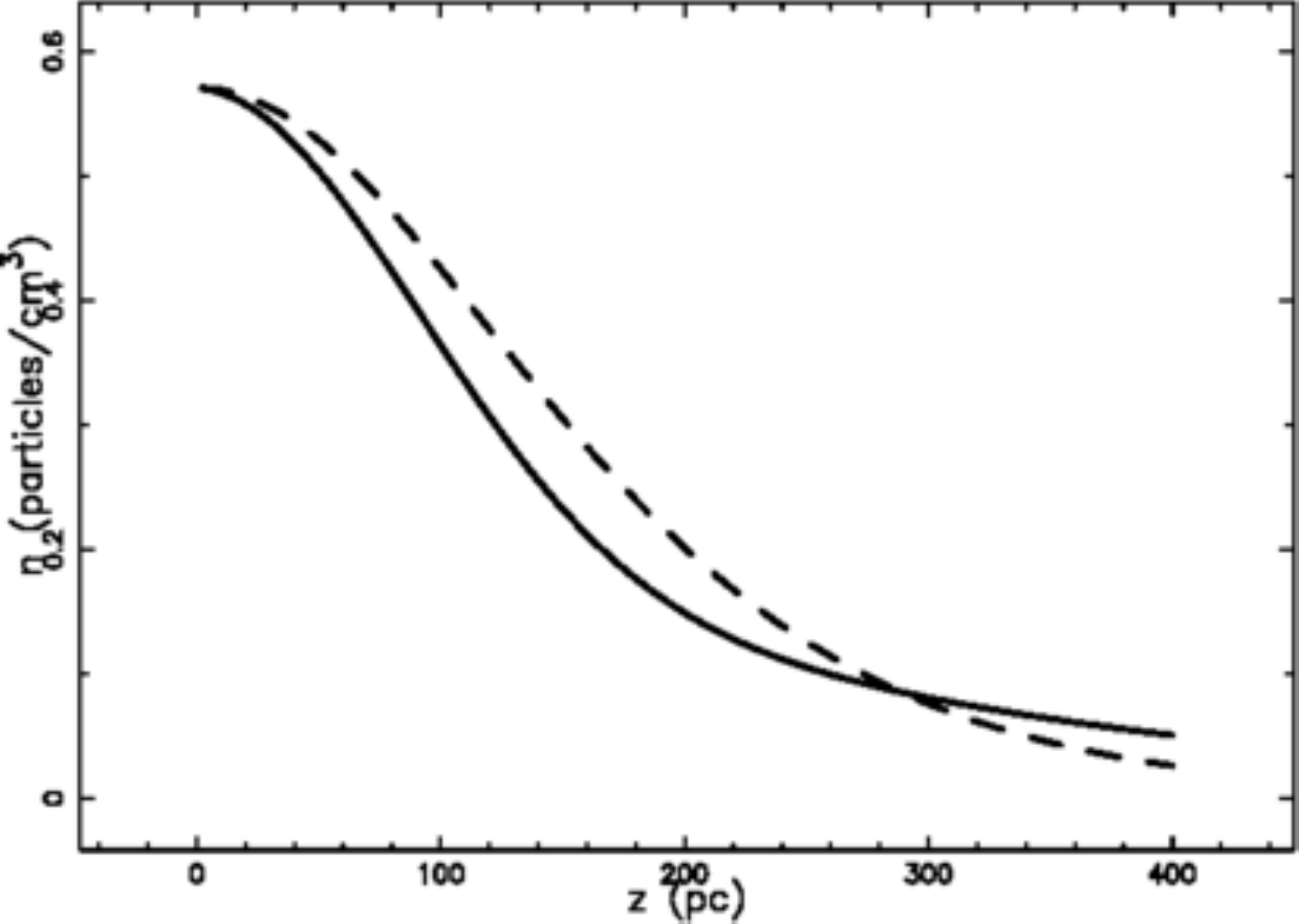}
  \end {center}
\caption
{
Profiles of density versus scale height $z$:
the  self-gravitating disk as
given by Eq.~(\ref{sech2})
when $h=90$\ pc
(dashed)
and
the
three-component exponential distribution
as
given by Eq.~(\ref{equation:ism})
(full line).
}%
    \label{zprofile}
    \end{figure}

Figure ~(\ref{zprofile})  compares 
the  empirical   function  sum of three exponential disks
and the theoretical  function
as given by Eq.~(\ref{sech2}).
% figure  zprofile
%end figure  zprofile
Assuming that the expansion starts  at 
$z=0$, we can write $z=R \sin (\theta)$, 
and therefore
\begin{equation}
n(R,\theta) = n_0 sech^2 (\frac{R \sin (\theta) }{2\,h})
\quad  ,
\label{sech2rtheta}
\end{equation}
where  $R$ is the radius of the advancing shell.

The 3D expansion that starts at the origin 
of the coordinates 
will be characterized by the following
properties.
\begin {itemize}
\item The dependence of the momentary radius of the shell
      on  the latitude  angle $\theta$ over the range
      $[-90 ^{\circ}  \leftrightarrow  +90 ^{\circ} ]$.

\item The independence of the momentary radius of the shell
      from  $\phi$, the azimuthal  angle  in the $x$-$y$  plane,
      which has a range
      $[0 ^{\circ}  \leftrightarrow  360 ^{\circ} ]$.
\end {itemize}
The mass swept, $M$,  along the solid angle
$ \Delta\;\Omega $  between 0 and $R$ is
\begin{equation}
M(R,\theta)=
\frac { \Delta\;\Omega } {3}  m_H n_0 I_m(R)
+ \frac{4}{3} \pi R_0^3 n_0 m_H
\quad  ,
\end {equation}
where
\begin{equation}
I_m(R)  = \int_{R_0} ^R r^2 
sech^2 (\frac{r \sin (\theta) }{2\,h})
dr
\quad ,
\end{equation}
where $R_0$ is the initial radius
and $m_H$ the mass of hydrogen.
The integral is
\begin{eqnarray}
I_m(R)  =  
-4\,h{r}^{2} ( \sin ( \theta )  ) ^{-1} ( 1
+{{\rm e}^{{\frac {r\sin ( \theta ) }{h}}}} ) ^{-1}+4
\,{\frac {h{r}^{2}}{\sin ( \theta ) }}
\nonumber \\
-8\,{h}^{2}r\ln 
 ( 1+{{\rm e}^{{\frac {r\sin ( \theta ) }{h}}}}
 )  ( \sin ( \theta )  ) ^{-2}
-8\,{h}^{3}{
\it 
\mathrm{Li}_{2}\/}
(-{{\rm e}^{{\frac {r\sin ( \theta ) }
{h}}}} )  ( \sin ( \theta )  ) ^{-3}
\quad ,
\end{eqnarray}
where 
\begin{equation}
\mathop{\mathrm{Li}_{2}\/}\nolimits\!\left(z\right)=\sum
_{{n=1}}^{\infty}\frac{z^{n}}{n^{2}}
\quad ,
\end{equation}
is the  dilogarithm,
see  \cite{Hill1828,Lewin1981,NIST2010}.

The conservation of momentum 
along the solid angle
$ \Delta\;\Omega $
gives
\begin{equation}
M(R,\theta)     \dot {R} (\theta) =
M(R_0)          \dot {R_0}
\quad  ,
\end{equation}
where $\dot {R}(\theta)$  is the  velocity
at $R$ and
$\dot {R_0}$  is the  initial velocity at $R=R_0$.
Using the previous equation, 
an analytical  expression  
for  $\dot {R}(\theta)$
along the solid angle
can be found, 
but it is complicated, and therefore we omit it.
In this differential equation of the first order in $R$, the
variables can be separated and integrating
term by term gives
\begin{equation}
\int_{R_0}^{R}  M(r,\theta)  dr =
M(R_0) \dot {R_0} \, ( t-t_0)
\quad  ,
\end{equation}
where  
$t$ is the time and 
$t_0$ the time at $R_0$.
We therefore have  an equation of the type 
\begin{equation}
{\mathcal{F}}(R,R_0,h)_{NL} =
\frac{1}{3} R_0^3
\dot {R_0} \, 
\left( t-{\it t_0}
 \right) 
\quad  ,
\label{fundamental}
\end{equation}
where  $  {\mathcal{F}}(R,R_0,h)_{NL}  $ has an analytical
but complicated form.
The  case  of expansion  that starts  from a given 
galactic height $z$,  denoted by $z_{\mathrm{OB}}$,
which  represent  the OB associations,  
cannot be solved by Eq. (\ref{fundamental}),  which 
is derived  for  a  symmetrical expansion that 
starts at  $z=0$.
It is not possible to find  $R$   analytically  and
a numerical method   should be implemented.
In  our case, in order
to find  the root of the nonlinear
Eq. (\ref{fundamental}), 
the FORTRAN subroutine  ZRIDDR from \cite{press} has been used.

The following two recursive equations are found when
momentum conservation is applied:
\begin{eqnarray}
R_{n+1} = R_n + V_n \Delta t    \nonumber  \\
V_{n+1} = V_n 
\Bigl (\frac {M_n(r_n)}{M_{n+1} (R_{n+1})} \Bigr ) 
\quad  ,
\label{recursive}
\end{eqnarray}
where  $R_n$, $V_n$, $M_n$ are the temporary  radius,
the velocity,  and the total mass, respectively,
$\Delta t $ is the time step,  and $n$ is the index.
The advancing expansion is computed in a 3D Cartesian
coordinate system ($x,y,z$)  with the center 
of the explosion at  (0,0,0).
The explosion is better visualized  
in a 3D Cartesian
coordinate system ($X,Y,Z$) in which the galactic plane
is given by $Z=0$.
The following 
translation, $T_{\mathrm{OB}}$,   
relates  the two Cartesian coordinate  systems. 
\begin{equation}
T_{\mathrm{OB}} ~
 \left\{ 
  \begin {array}{l} 
  X=x  \\\noalign{\medskip}
  Y=y  \\\noalign{\medskip}
  Z=z+ z_{\mathrm{OB}}
  \end {array} 
  \right.  \quad , 
\label{ttranslation}
\end{equation}
where $z_{\mathrm{OB}}$  
is the distance  in parsecs  of the 
OB associations   from the galactic plane.

The physical units have not yet been specified: 
parsecs for length and
$10^7\,yr$ for time,
see also subsection \ref{astrounits},
are an acceptable 
astrophysical choice.
With
these units, the initial velocity $V_{{0}}=\dot {R_0}$ is 
expressed in
units of pc/($10^7$ yr) and should be converted 
into km/s; this
means that $V_{{0}} =10.207 V_{{1}}$ 
where  $V_{{1}}$ is
the initial velocity expressed in km/s.

Analytical  results  can also be obtained 
solving  the Kompaneets
equation, see \cite{Kompaneets1960}, 
for the motion
of a shock wave in different plane-parallel 
stratified media   such as 
exponential, power-law type, and a 
quadratic hyperbolic-secant, 
see synoptic Table  4 in   \cite{Olano2009}.

\section{A Comparison with Hydrosimulations} 
\label{sectionhydro}
This Section reports the comparison 
of the thermal and cold models 
with numerical hydro-dynamics 
calculations.
\subsection{The thermal test}

The shape   of the SB in the thermal model,
see Section \ref{sectionthermal},  
depends   strongly   on
the time
 elapsed since the \underline{first} explosion, 
$t_{\mathrm {age}}$,
the \underline{duration of SN burst},
$t^{\mathrm{burst}}$,
 on
the \underline{number}
of SN explosions
in the bursting \underline{phase},
and on the adopted density profile.
Ordinarily, the dynamics of an SB is studied
with hydro simulations
(partial differential equations).
 Conversely, here under the hypothesis
called ``thin-shell approximation'', the dynamics is studied
by solving ordinary differential equations.
This choice  allows the calculation
of  a much larger number of models compared
with hydro-dynamics  calculations.

The level  of confidence in   our results
can  be given by a comparison with numerical hydro-dynamics 
calculations  see, for example, 
\cite{MacLow1989}.
The vertical density distribution
they  adopted $[$ see equation(1) from 
\cite{MacLow1989} and
equation(5) from \cite{Tomisaka1986} $]$
has  the following
{it z} dependence, 
the  distance  from  the galactic plane in pc:
\begin {equation}
n_{\mathrm {hydro}} =
n_{\mathrm {d}} \left \{
\Theta {\mathrm {exp}} [-\frac {V(z)}{\sigma_{{\mathrm {IC}}}^2} ]
+(1-\Theta) {\mathrm {exp}}
\left  [-\frac {V(z)}{\sigma_{{\mathrm {{\mathrm {C}}}}}^2} \right ]
\right\}
,
\label{ISMhydro}
\end {equation}
with the gravitational potential as
\begin{equation}
V(z)=
68.6 {\mathrm {ln}} \left 
[1 + 0.9565~{\mathrm { sinh}}^2 \left (0.758 \frac {z}{z_0}\right) \right ]
({\mathrm {km s}}^{-1})^2
.
\end {equation}
Here $n{\mathrm {_d}}=1~\mbox{particles~cm}^{-3} $, $\Theta=0.22$,
$\sigma_{IC}=14.4{\mathrm { km~s}}^{-1}$,
$\sigma_{C}=7.1  {\mathrm { km~s}}^{-1}$
and $z_0 =124~{\mathrm {pc}}$.
 Table \ref{tab:rel} reports the results of ZEUS
(see \cite{MacLow1989}),  a two-dimensional hydrodynamic   code, when
$t_{\mathrm {age}}=0.45 \cdot 10^7~{\mathrm{yr}}$.  
The supernova luminosity is
$1.6 \cdot 10^{38} {\mathrm{erg~s}}^{-1}$, $z_{\mathrm{OB}}$=100~pc
and
the density distribution is given by
formula~(\ref{ISMhydro}).
The ZEUS code was  originally described by~
\cite{Stone1992}.

\begin{figure}
  \begin{center}
\includegraphics[width=7cm]{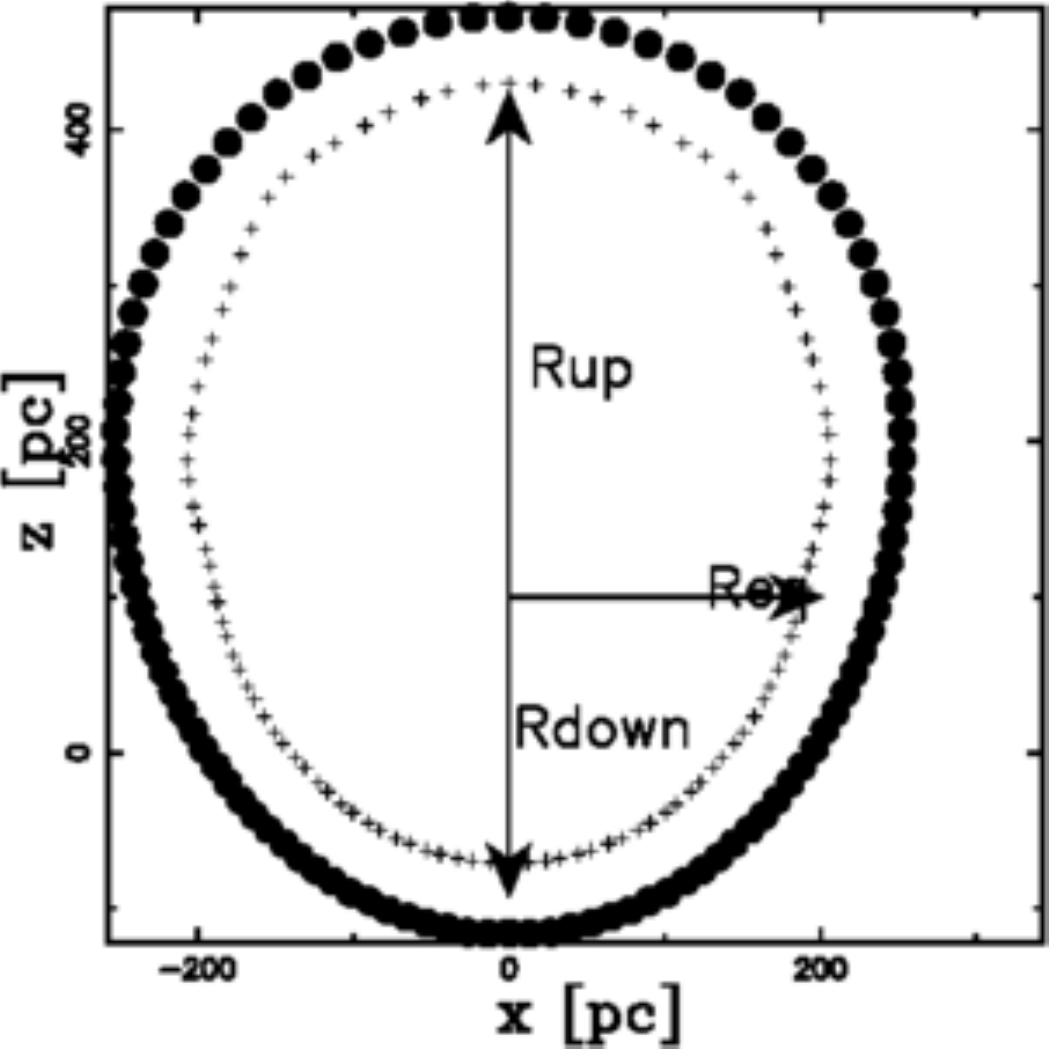}
  \end{center}
\caption
{
Section of the  on the {\it x-z}  plane
when the explosion starts at  $z_{\mathrm{OB}}=100 \mbox{pc} $.
The density law is given by
Equation \ref{ISMhydro}.
The code parameters are
$t_\mathrm{\mathrm {age}}$=$0.45~10^{7}~\mathrm{yr}$,
$\Delta t=0.001\cdot10^{7}~\mathrm{yr}$,
$t^{\mathrm{burst}}_7$=0.45,
  and  $N^*$=248.
The points represented by the  small crosses indicate 
the inner section from  Figure 3a  of \cite{MacLow1989}.
}
\label{confronto}%
    \end{figure}

In order to make a comparison  our code  was run
with the parameters of the hydro-code
(see Figure \ref{confronto});
a density profile as  given by
formula~(\ref{ISMhydro}) was adopted.
The percentage  of
reliability  of our code can also be  introduced,
\begin{equation}
\epsilon  =(1- \frac{\vert( R_{\mathrm {hydro}}- R_{\mathrm {num}}) \vert}
{R_{\mathrm {hydro}}}) \cdot 100
\,,
\label{efficiency}
\end{equation}
where $R_{\mathrm {hydro}}$ is the   radius,
as given by the hydro-dynamics,
and  $R_{\mathrm {num}}$ the radius  obtained from our  simulation.
In the already cited table \ref{tab:rel},
our numerical radii can also be found
in the    upward, downward and  equatorial
directions,
and the efficiency  as  given by
formula~(\ref{efficiency}).
The value of the radii are comparable  in
all of the three  chosen directions and
Figure \ref{confronto} also reports on   the data
from  Figure 3a  of \cite{MacLow1989}.

\begin{table}
      \caption{Code reliability. }
         \label{tab:rel}
      \[
         \begin{array}{ccccc}
            \hline
            \noalign{\smallskip}
~~~~     & R_{\mathrm{up}}(\mathrm{pc})  &
         R_{\mathrm{down}}(\mathrm{pc})  &
         R_{\mathrm{eq}}  (\mathrm{pc})  \\
            \noalign{\smallskip}
            \hline
            \noalign{\smallskip}
R_{\mathrm {hydro}} (ZEUS)     &  330  &  176   &  198        \\
R_{\mathrm {num}}   (\mbox{our~code}) &  395  &  207  &  237
        \\
\mbox {efficiency} (\%)      &  80  &  81  &   80        \\
            \noalign{\smallskip}
            \hline
         \end{array}
      \]
   \end{table}

\subsection{The cold  test}

\begin{figure}
  \begin{center}
\includegraphics[width=7cm]{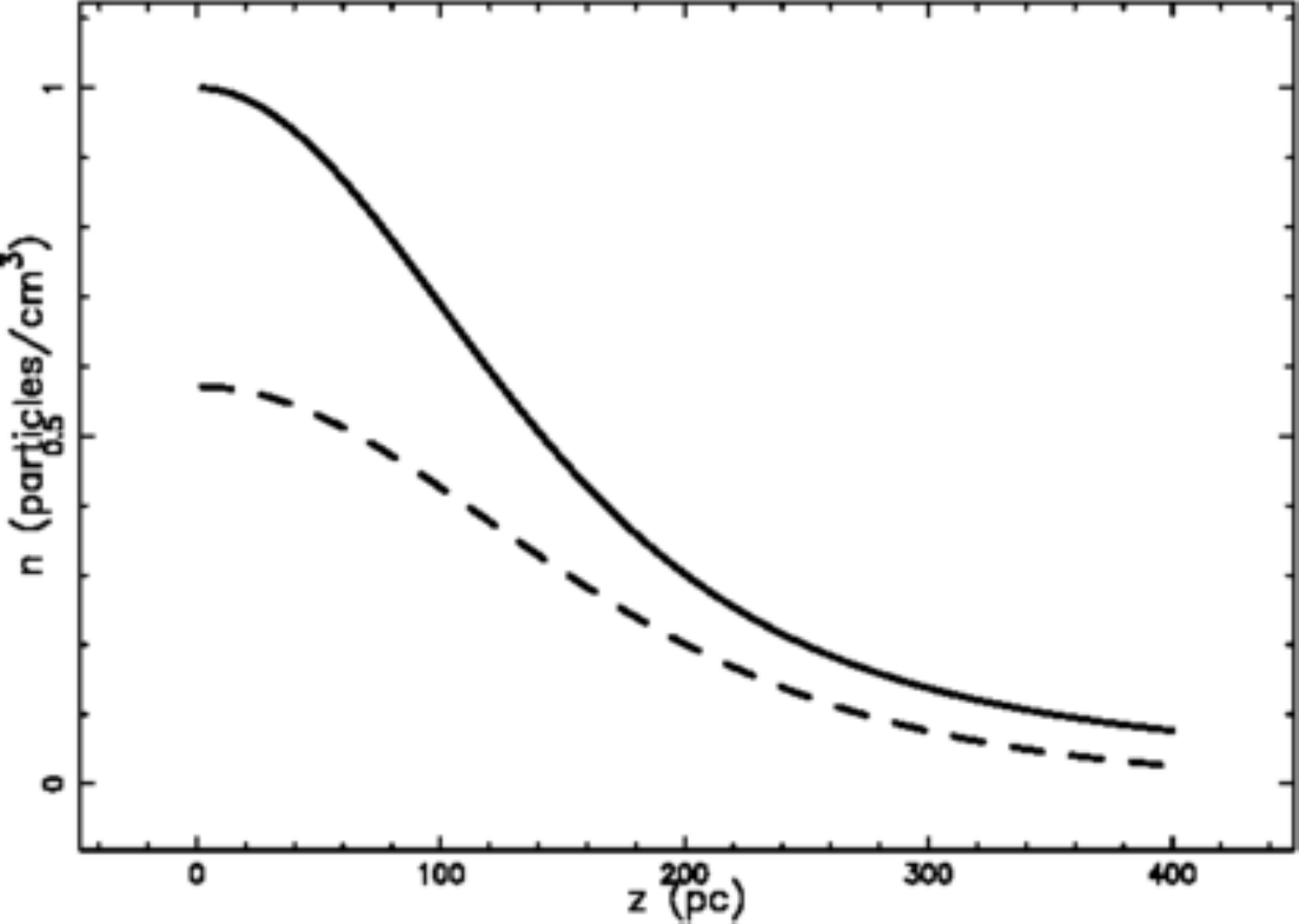}
  \end {center}
\caption
{
Profiles of density versus scale height $z$:
the  self-gravitating disk as
in Eq.~(\ref{sech2})
when $h=90$\ pc
(dashed line)
and the hydro number density as given by Eq. \ref{ISMhydro}
(full line).
}%
    \label{zprofilehydro}
    \end{figure}

In the framework  of the cold model 
developed in Section  \ref{sectioncold} 
Figure  \ref{zprofilehydro}  compares
the  hydro number density as given by Eq. \ref{ISMhydro}
and the theoretical  function
as given by  Eq.~(\ref{sech2}).
The difference in the density profiles from hydrosimulations and 
from the cold model adopted 
are due to the fact  that the density 
at  $z=0$ is  assumed to be   
$n_{\mathrm {hydro}}=1~\mbox{particles~cm}^{-3} $
in the hydrosimulations, see \cite{MacLow1989}.
In our  cold model conversely  at  $z=0$ we have  
$n=0.566~\mbox{particles~cm}^{-3} $
as  in \cite{Lockman1984}.
% figure  zprofilehydro
%end figure  zprofilehydro

\begin{figure}
\includegraphics[width=7cm]{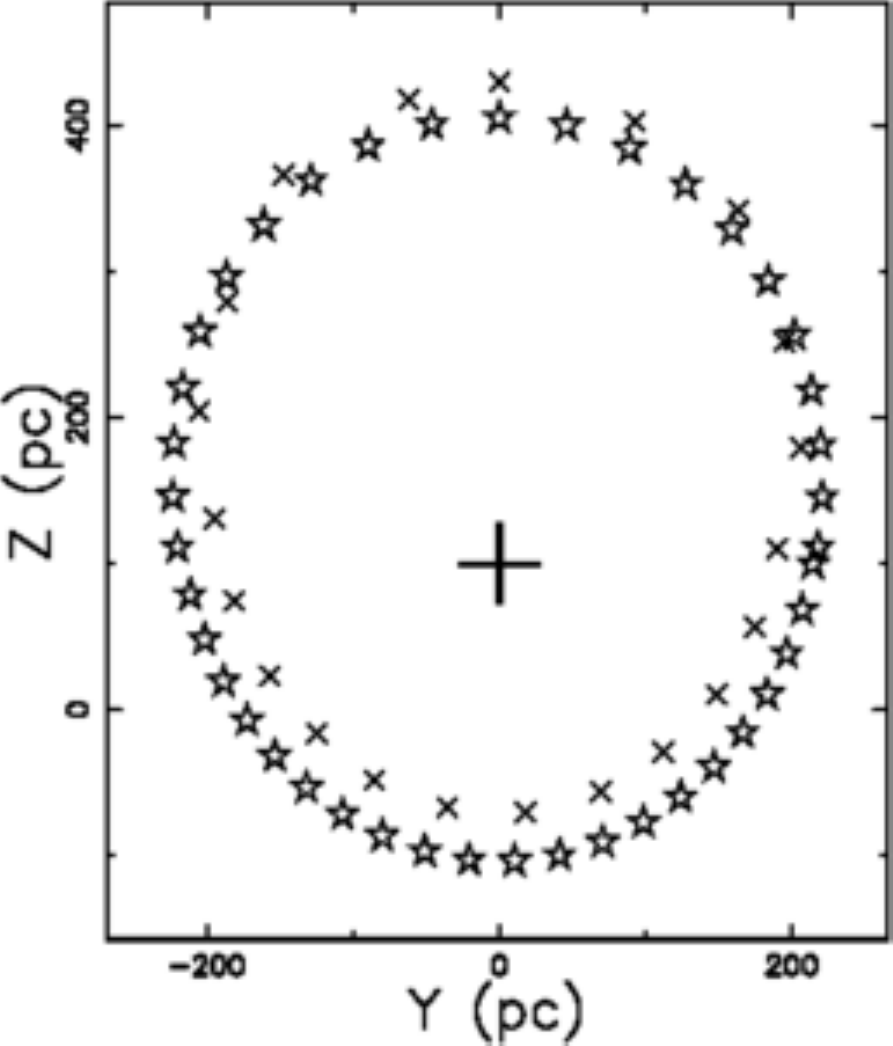}
\caption
{
Section of the SB in the         {\it X-Z}  plane
when the explosion starts at  $z_{\mathrm{OB}}=100 \mbox{pc} $
(empty stars).
The cold code parameters are
$h=90$ pc, 
$t_7=0.45~$,
$t_{7,0}$ = 0.00045,
$r_0$  = 24.43, 
$V_0= 3191\, \mathrm{km}\, \mathrm{s}^{-1}$,
$N_{SN}$ = 180   
 and  $N^*$=2000000.
The points represented by the  small crosses indicate 
the inner section from  Figure  3a  of  MacLow et al. 1989.
The explosion site is represented by a big cross.
}
\label{autoccc}%
    \end{figure}

A typical run  of ZEUS
(see \cite{MacLow1989}),  a two-dimensional hydrodynamic   code, 
is  done for 
$t_7 =0.45 $ and   
 supernova luminosity of 
$1.6 \cdot 10^{38} {\mathrm{erg~s}}^{-1}$
when  $z_{\mathrm{OB}}$=100~pc.
In order to make a comparison with  our
cold code, we  adopt the same time  
and we search the  parameters 
which  produce similar  results, see  
Figure  \ref{autoccc}.
%inizio figure autoccc
%fine figure autoccc

\section{Astrophysical SBs}
\label{sectionastro}

This section  evaluates the significance of the
galactic rotation  for the shape of the SBs,
models 
the thermal GW~46.4+5.5, 
the thermal Gould Belt,
the cold GW~46.4+5.5
and    
the thermal  Galactic Plane. 
\subsection{Galactic rotation}

The  influence of the Galactic rotation on the results 
can be be obtained by  introducing 
the  law  of the Galactic rotation 
as given by \cite{Wouterloot_1990},
\begin{equation}
V_{\mathrm{R}} (R_0)  =220 ( \frac {R_0[\mathrm{pc}]} {8500})^{0.382} 
\mathrm {km~sec}^{-1}
,
\label {vrotation}
\end {equation} 
where $R_0$ is the radial distance from the center of the Galaxy 
expressed in pc.
The translation of the previous formula to 
the astrophysical units  adopted gives 
\begin{equation}
V_{\mathrm{R}} (R_0)  =2244 \left ( \frac {R_0[\mathrm{pc}]} 
{8500}\right )^{0.382} \frac {\mathrm{pc}}{10^7~{\mathrm{yr}}}
,
\label {vrotationpc}
\end {equation}
\begin{equation}
\Omega  (R_0)  =2244 \frac {( \frac {R_0[\mathrm{pc}]} {8500})^{0.382}} 
{R_0[\mathrm{pc}]}  \frac {\mathrm{rad}}{10^7~{\mathrm{yr}}}
.
\label {omegadiff}
\end {equation}
Here, $ \Omega (R_0)  $ is the  differential angular velocity 
and  
\begin{equation}
\phi   =2244 \frac {( \frac {R_0[\mathrm{pc}]} {8500})^{0.382}}
{R_0[\mathrm{pc}]}    t_7    ~~~\mathrm{rad}.
\label {phitotale}
\end {equation}
Here, $  \phi   $, is the angle 
made on the circle and $t_7$ the time expressed
in units of $10^7~{\mathrm{yr}}$. 
Upon  considering only a  single object,
  an  expression for the
angle  can be found  once  $R=R_0+y$ is introduced, 
\begin{equation}
\phi(y)  =  \frac  {V_{\mathrm {R}} (R_0)}{R_0+y}  t. 
\label{fi1} 
\end {equation}
The shift in the angle due to  differential rotation
can now be introduced,
\begin{equation}
\Delta \phi   = \phi(y) - \phi(0) 
,
\label{fi2}
\end {equation}
where {\it x } and {\it y} are the SB coordinates in the 
inertial frame 
of  the explosion, denoted by  {\it x}=0,{\it y}=0 and 
{\it z}=$z_{\mathrm{OB}}$.
The great distance from the center allows us to say
that   
\begin{equation}
\frac {\Delta x}{R_0} 
=   \Delta \phi 
,
\label {deltax}
\end{equation}
where  $\Delta x $ is the shift due to the differential rotation in
the {\it x} coordinate.
The shift  
can be found from (\ref{deltax}) and (\ref{fi2})
once  a Taylor  expansion is performed,  
\begin{equation}
\Delta x  \approx -  V_{\mathrm {R}} (R_0)\frac {y}{R_0} t
 .
\label {xt}
\end{equation}
Upon inserting (\ref{vrotationpc}) in (\ref{xt}),
 the following 
transformation, $T_{\mathrm {r}}$,   due to the rotation is
obtained for a single object in the solar surroundings:
\begin{equation}
T_{\mathrm{r}} ~
 \left\{ 
  \begin {array}{l} 
  x\prime=x  +  0.264 y ~t\\\noalign{\medskip}
  y\prime=y\\\noalign{\medskip}
  z\prime=z,
  \end {array} 
  \right. 
\label{trotation}
\end{equation}
where {\it y} is expressed in pc and $t$ in units of $10^7~{\mathrm{yr}}$.

\subsection{Thermal GW~46.4+5.5}

\label{secthermalgw46}
A careful study of the worms 46.4+5.5 and  39.7+5.7 \cite{Kim2000}
has led to the conclusion that they  belong to a single
super-shell. Further on,  the dynamical properties  of this H\,I 
supershell can be deduced by coupling the observations with 
theoretical arguments,  \cite{Igumenshchev1992}; 
the derived  model parameters to fit the observations 
are reported in  Table \ref{tab:ssh},
where the  altitudes of 
KK~99~3 and KK~99~4   (clouds that are CO emitters)
have been identified with $z_{\mathrm{OB}}$ by the author.

   \begin{table}
      \caption{Data of the supershell associated with GW~46.4+5.5.}
         \label{tab:ssh}
      \[
         \begin{array}{cc}
            \hline
            \noalign{\smallskip}
\mbox {Size~(pc}^2)                    & 345 \cdot 540  \\
\mbox {Expansion~velocity~ (km~s$^{-1}$}) & 15              \\
\mbox {Age~(10$^7$~yr})                    & 0.5             \\
\mbox {z$_{OB}$  (pc)}                   & 100             \\
\mbox {Total~energy~($10^{51}$}{\mathrm{erg}})      & 15              \\
            \noalign{\smallskip}
            \hline
         \end{array}
      \]
   \end{table}
 
\begin{figure}
\includegraphics[width=7cm]{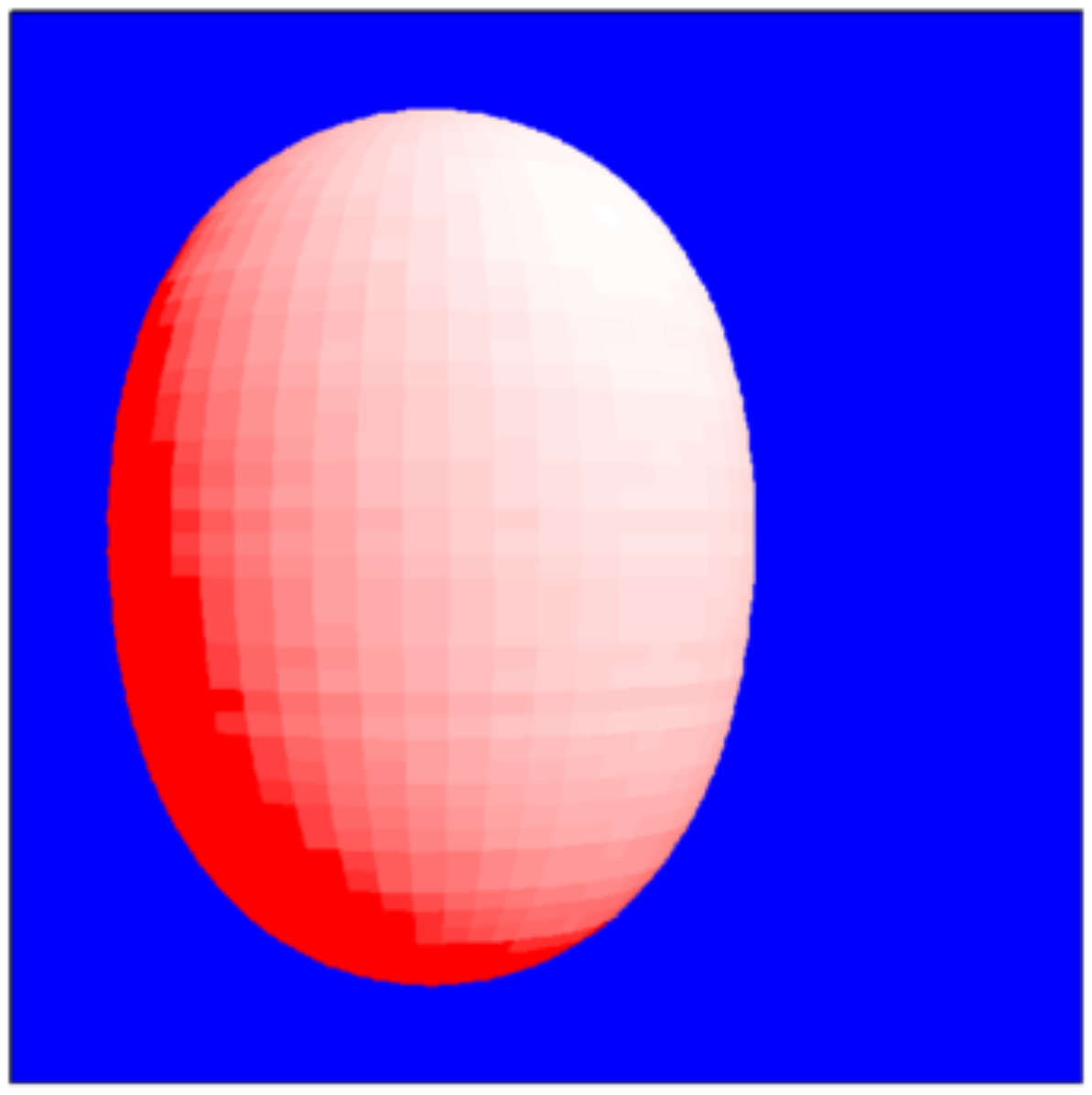}
\caption
{
Model of  GW~46.4+5.5.
The parameters are
$t_\mathrm{\mathrm {age}}$=$0.5 \cdot 10^{7}~\mathrm{yr}$,  
$\Delta t=0.001 \cdot 10^{7}~\mathrm{yr}$,
$t^{\mathrm{burst}}_7$= 0.5, $N^*$=  150, $z_{\mathrm{OB}}$=100 pc, 
and $E_{51}$=1.
The three Eulerian angles 
characterizing the point of view of the observer
are  $ \Phi $= 0$^{\circ }$
, $ \Theta $= 90$^{\circ }$
 and  $ \Psi $= 0$^{\circ }$.
}
\label{worm}%
    \end{figure}

These parameters   are  the input for  our thermal computer 
code (see the captions of Figure \ref{worm}).
The problem of assigning a value to $z_{\mathrm{OB}}$ now arises,
 and the
following two equations are set up:
\begin{eqnarray}
R_{\mathrm{up}}  +z_{\mathrm{OB}}=& 540&~  , \nonumber   \\
\frac{R_{\mathrm{up}}+z_{\mathrm{OB}}}{R_{\mathrm{down}}-z_{\mathrm{OB}}}=& \frac{D_{\mathrm{up}}}{D_{\mathrm{down}}}= &
\frac{15^{\circ}}{3^{\circ}}=3.
\label{system}
\end {eqnarray}
The algebraic system~(\ref{system}) consists of two equations and
three variables. The value chosen for the minimum and maximum
latitudes ($15^{\circ}$    and   $-5^{\circ}$)
 are in rough agreement with the position of the center at
$+5^{\circ}$ of the galactic latitude (see \cite {koo}). 
One way 
of solving 
\ref{system} is to set, for example, $z_{\mathrm{OB}}$=100~pc.
The other two variables are easily found to be as  follows:
\begin{eqnarray}
z_{\mathrm{OB}}   =& 100~\mbox {pc}  ,\nonumber   \\
R_{\mathrm{down}} =& 235~\mbox {pc}  ,           \\
R_{\mathrm{up}}  =& 305~\mbox  {pc}
 \nonumber  .
\label{solution}
\end {eqnarray}
For this value of 
  $z_{\mathrm{OB}} $,   we 
have  the case where a transition from an egg-shape to a V-shape
is going on,    and the simulation gives the exact shape.    
In order to obtain 
$E_{\mathrm{tot}}=15. \cdot 10^{51}{\mathrm{erg}}$ 
(see formula~(\ref{etotburst}))
and  $t^{\mathrm{burst}}=0.5 \cdot 10^7~{\mathrm{yr}}$,
we have inserted 
$N^*$=~150.

In order to  test our simulation,  an
observational percentage  of 
reliability is  introduced that uses 
both the size and the shape, 
\begin{equation}
\epsilon_{\mathrm {obs}}=100(1-\frac{\sum_j |R_{\mathrm {obs}}-R_{\mathrm {num}}|_j}{\sum_j
{R_{\mathrm {obs}}}_j})
, 
\label{eq:reliability}  
\end{equation}
where $R_{\mathrm {obs}}$ is the observed  radius,
as deduced  by 
using the following algorithm. 
The radius at regular angles from the  vertical 
($0{^{\circ}}, ~90{^{\circ}}, ~180{^{\circ}}$) is extracted from
Table \ref{tab:ssh},
 giving the series (305~pc, ~172.5~pc, ~235~pc);
the theory of cubic splines \cite{press}  
is then applied to compute
the various radii at 
progressive angles 
($0{^{\circ}}, ~3.6{^{\circ}}, ~7.2{^{\circ}},...180{^{\circ}}$),
which  is a series computed by adding 
regular steps of  $180~{^{\circ}}/n_{\theta}$. 
The data are extracted from the dotted ellipse
visible in Figure 7 of \cite{koo};
this  ellipse represents the super-shell at 
$v_{LSR}=18.5~\mbox{km\,sec}^{-1}$.

\begin{figure}
\includegraphics[width=7cm]{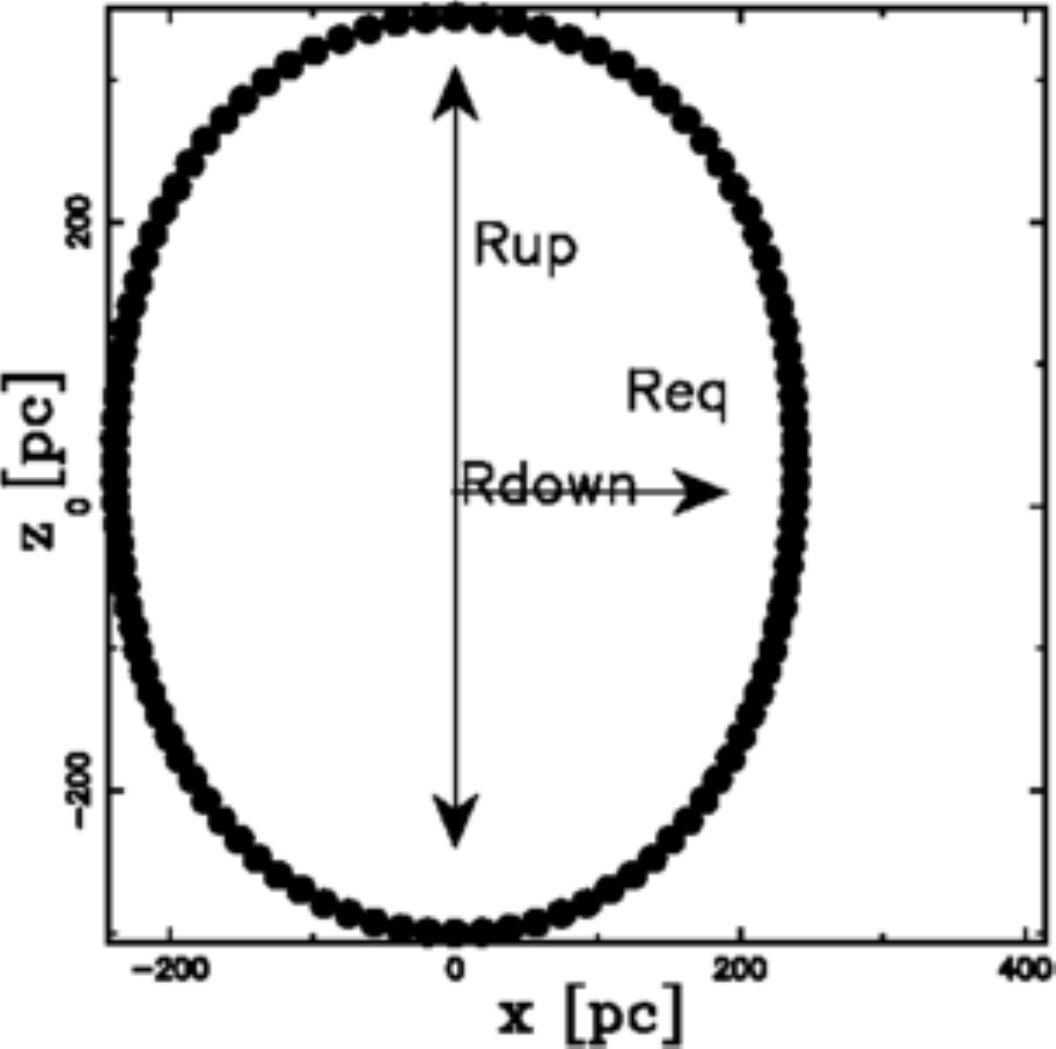}
\caption
{Section of the SB on the {\it x-z }
 plane    when the   
physical parameters are the  same as in Figure \ref{worm}.
}
\label{worm_sect}%
    \end{figure}

We can now compute the efficiency over 50+1 directions
$[$ formula~(\ref{eq:reliability})$]$  of a section {\it y-z} when
 {\it x}=0  
which  turns out to be 
$\epsilon_{\mathrm {obs}}=68.4 \%$;  
the observed and numerical
radii along  the three typical directions are reported
in Table \ref{tab:rel_gw}.
The physical parameters adopted from \cite{Kim2000}  
turn out to be consistent with our thermal numerical code.
   \begin{table}
      \caption{Radii concerning  GW~46.4+5.5.}
         \label{tab:rel_gw}
      \[
         \begin{array}{cccc}
            \hline
            \noalign{\smallskip}
\mathrm{Direction}& R_{\mathrm {num}}(\mathrm{pc}) &R_{\mathrm
{num}}(\mathrm{pc})
\mbox { with
 the Euler process}& R_{\mathrm {obs}}(\mathrm{pc}) \\
            \noalign{\smallskip}
            \hline
            \noalign{\smallskip}
\mathrm{Equatorial} &  238  & 233 &   172.5   \\
\mathrm{Polar~up}   &  342  & 335 &   305     \\
\mathrm{Polar~down} &  312  & 304 &   235     \\
            \noalign{\smallskip}
            \hline
         \end{array}
      \]
   \end{table}
The results of the simulation can be visualized in Figure \ref{worm},
or through a  section on the {\it x-z}
plane  (see 
Figure \ref{worm_sect}).
\label{ref_velocity}
Another  important  observational  parameter is the H\,I
21~cm line emission: the observation of H\,I gas associated 
with GW~46.4+5 \cite{Hartmann1997,Kim2000}  
reveals that the super-shell has 
an expansion velocity 
of  $V_{\mathrm{exp}} \approx$ 15  $\mbox{km\,s}^{-1}$ \cite{Kim2000}.
In order to see how our model matches  the observations,
the instantaneous radial velocities 
are computed in each direction.
A certain number of random points are then generated 
on the theoretical surface of expansion. The relative velocity
of each point is  computed
by using the method  of   bilinear interpolation on the four 
grid points that surround the selected latitude and 
longitude \cite{press}. \label{para:velo}
 
\begin{figure}
\includegraphics[width=7cm]{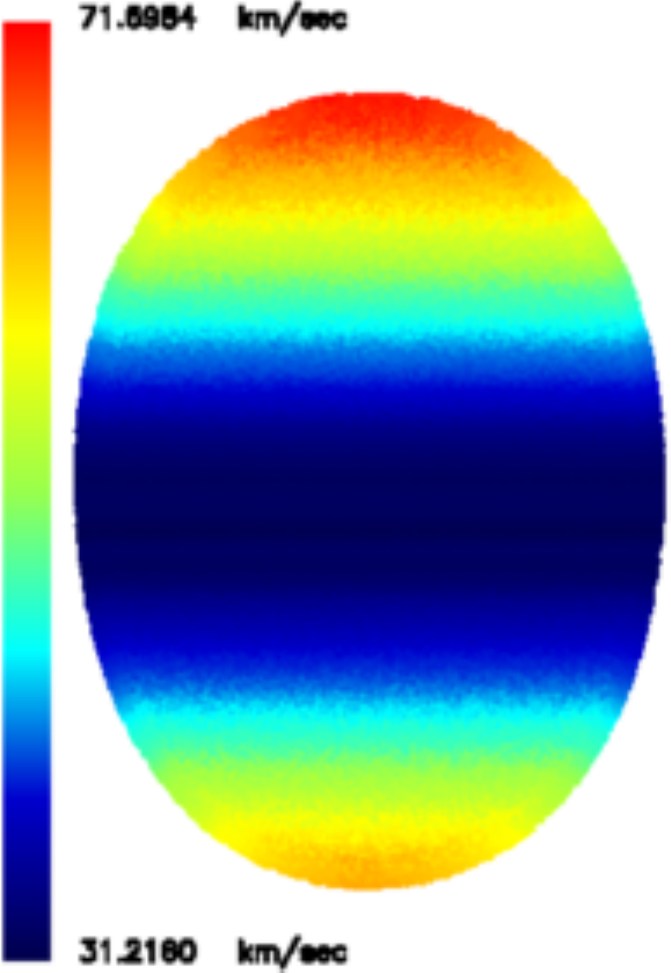}
\caption
{
  Map of the expansion velocity  
  relative to a simulation of GW~46.4+5.5~ 
  when 190000 random points are selected on the surface.
  The   physical parameters are the  same as in Figure \ref{worm}
  and 
  the three Eulerian angles 
  characterizing the point of view of the observer
  are  $ \Phi $= 0   $^{\circ }$
, $ \Theta $= 90   $^{\circ }$,
  and  $ \Psi $=0   $^{\circ }$.
}
\label{worm_velocita}%
    \end{figure}

Our model gives radial velocities of $V_{\mathrm{theo}}$, 
31~ $\mbox{km~ s}^{-1}$ $\leq~V_{\mathrm{theo}}~\leq$ 71 ~
$\mbox{km~s}^{-1}$
(27~ $\mbox{km~s}^{-1}$ $\leq~V_{\mathrm{theo}}~\leq$ 66 ~
$\mathrm{km~s}^{-1}$
with the Euler process)
and a  map of the expansion velocity 
is given in Figure \ref{worm_velocita},
from which it is possible to visualize the differences 
in the expansion velocities between the various regions.

\begin{figure}
\includegraphics[width=7cm]{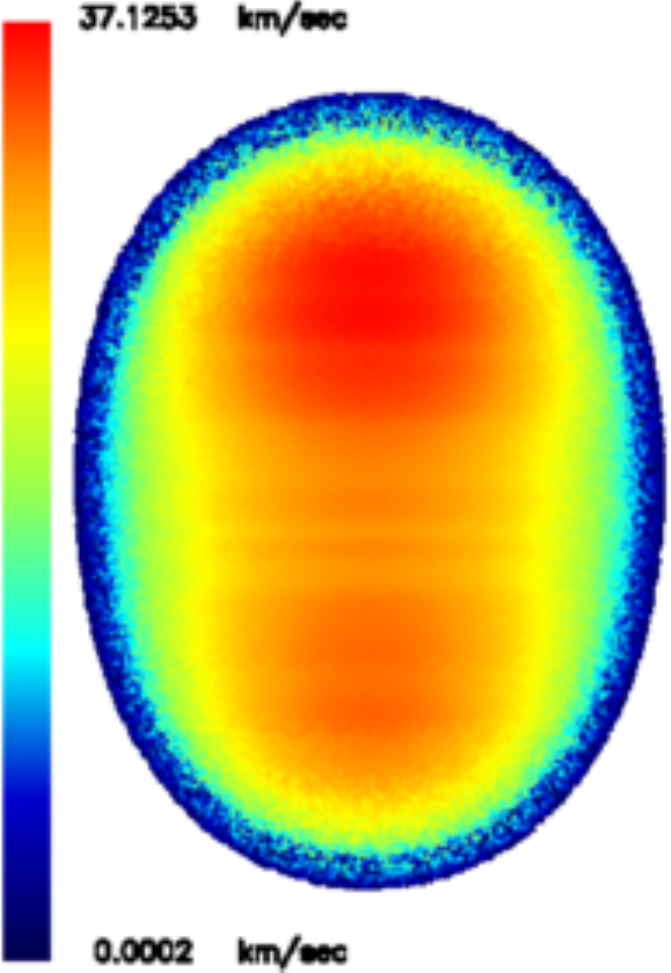}
\caption
{
 Map of the velocity along the line of sight, 
 $V^{\mathrm p}_{\mathrm {theo}}$, 
  relative to a simulation of GW~46.4+5.5~ 
  when 190000 random points are selected on the surface.
  The   physical parameters are the  same as in 
  Figure \ref{worm_velocita_y}.
}
\label{worm_velocita_y}%
    \end{figure}

Perhaps it  is useful  to map the velocity ($V^{\mathrm p}_{\mathrm {theo}}$)
 in the  {\it y}-direction,  
\begin {equation}
V^{\mathrm p}_{\mathrm {theo}} = v (\theta,\phi) \cdot  \sin (\theta) \cdot (\sin \, \phi).
\end   {equation}
This is the velocity measured along the line  of sight 
when an observer stands in front of the super-bubble;
$\theta$  and $\phi$  were defined in subsection \ref{sectheta}.
The  structure of the projected velocity, $V^{\mathrm p}_{\mathrm{theo}}$, 
is  mapped
in Figure \ref{worm_velocita_y}~
by using different colors; 
the range  is 0~$\mbox{km~s}^{-1}$ 
$\leq~V^{\mathrm p}_{\mathrm {theo}}~\leq$ 36~$\mbox{km~s}^{-1}$
(0~$\mbox{km~s}^{-1}$ 
$\leq~V^{\mathrm {p}}_{\mathrm {theo}}~\leq$ 32~$\mbox {km~s}^{-1}$
with the Euler process)
and  the averaged  projected velocity  is  
$\approx 20 \, \mbox {km~s}^{-1}$ 
($\approx 17.6 \, \mbox {km~s}^{-1}$ with the Euler process),
a value  that is greater by  $\approx 5\,\mbox {km~s}^{-1}$ 
($\approx 2.6 \mbox {km~s}^{-1}$ with the Euler process) than the already
mentioned  observed expansion velocity,
$V_{\mathrm {exp}}=15\,\mbox{km~s}^{-1}$.  

As is evident from the map in   Figure \ref{worm_velocita_y},
the projected velocity of expansion is not uniform
over all of the shell's surface,
  but is greater in the central
region than in the external region.
In this particular case of an  egg-shape,
we observe 
a  nearly circular region connected 
with the maximum velocities in the upper part of the shell.
It is  therefore possible to speak of egg-shaped 
appearances 
in the 
Cartesian  physical coordinates  
and  spherical-appearances in the projected maximum velocity.

\subsection{Thermal Gould Belt}

\label{gouldbelt}
\begin{figure}
\includegraphics[width=7cm]{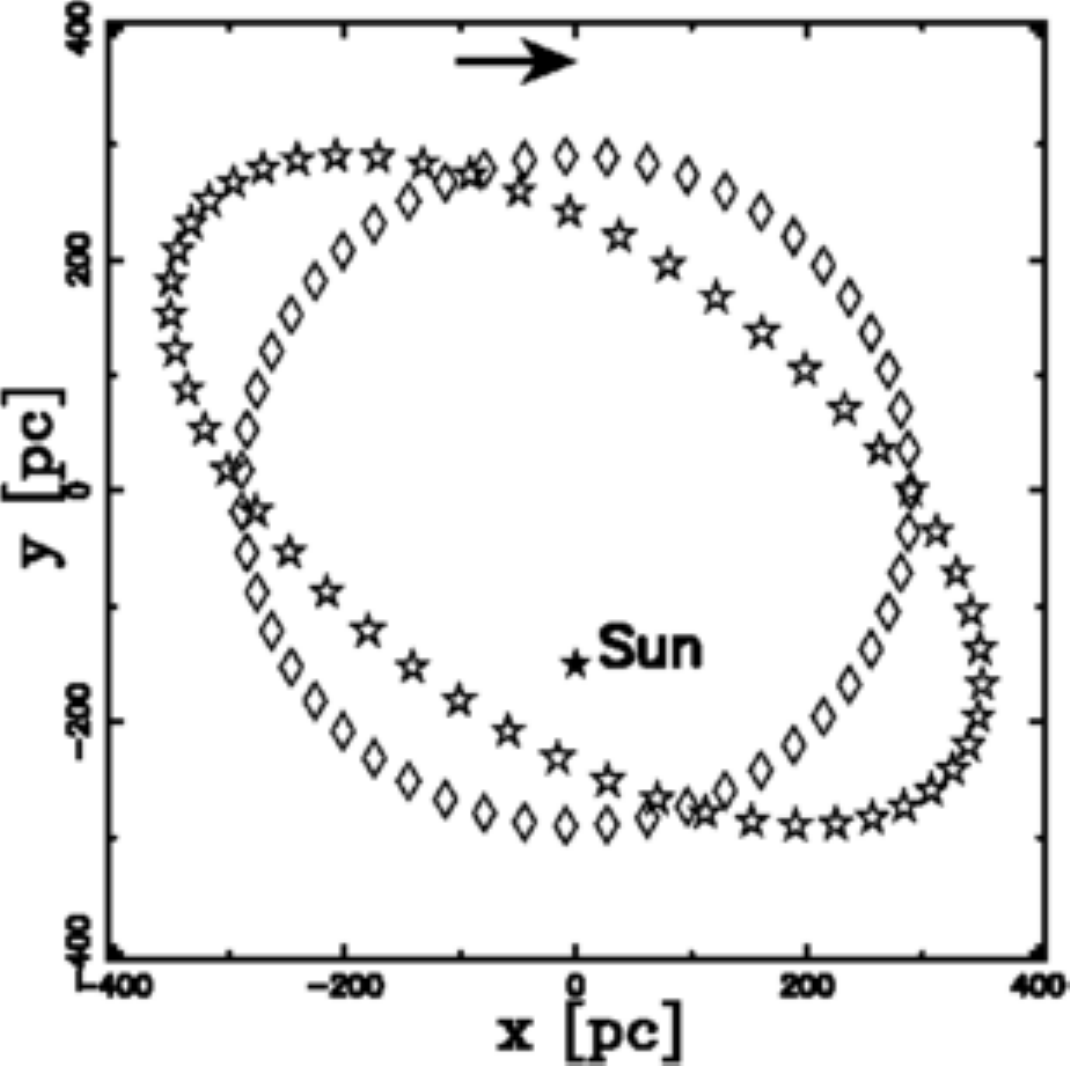}
\caption
{
Rhombi represent the circular  section,
the stars 
the rotation-distorted section and the big star the Sun.
The Galaxy's direction of  rotation is also  shown.  
The parameters are
$t_\mathrm{\mathrm {age}}$=$2.6\cdot10^{7}~\mathrm{yr}$,  
$\Delta$~$t=0.001\cdot10^{7}~\mathrm{yr}$,
$t^{\mathrm{burst}}_7$= 0.015, $N^*$=  2000, $z_{\mathrm{OB}}$=0 pc,
and $E_{51}$=1.
}
\label{rotation_gould_sun}%
\end{figure}

\begin{figure}
\includegraphics[width=7cm]{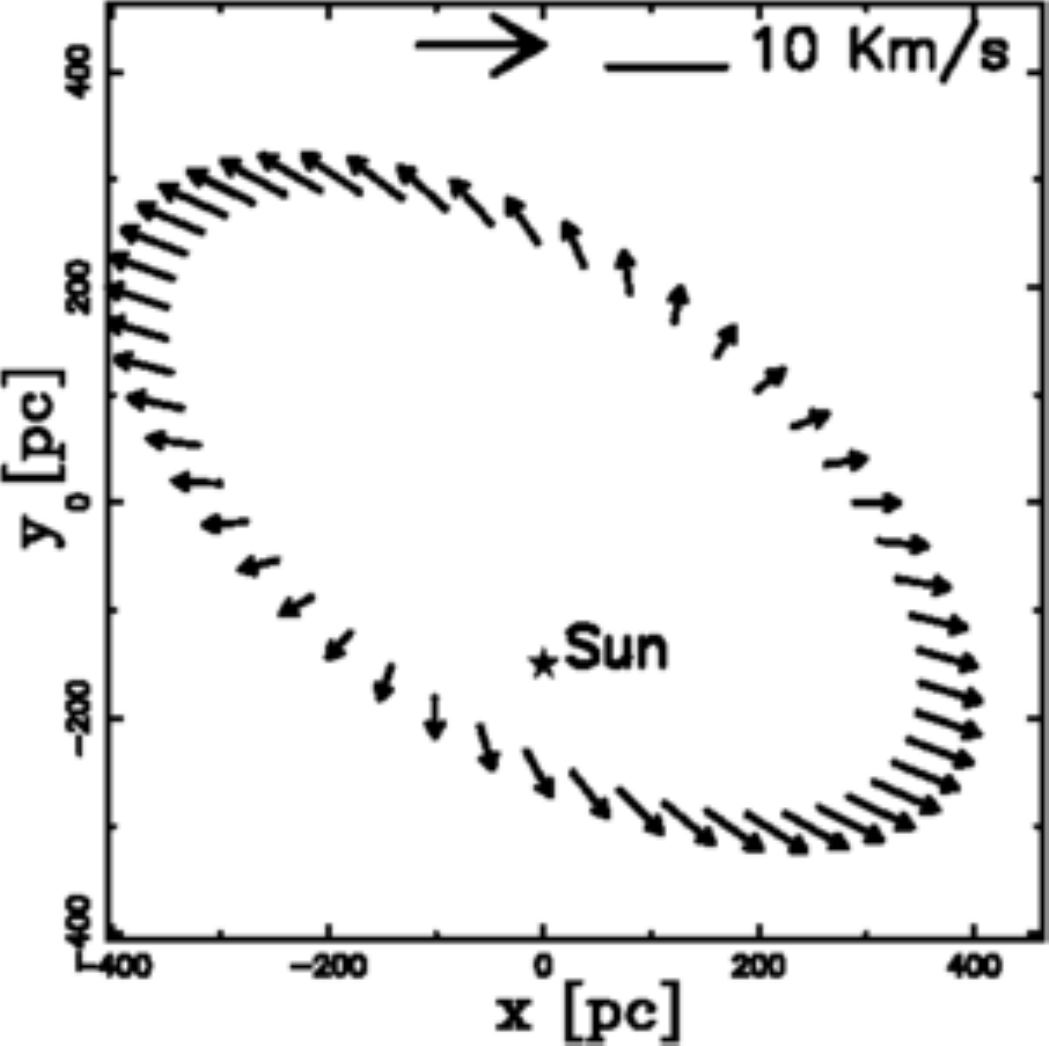}
\caption
{
The stars represent 
the rotation-distorted section of the Gould Belt  and the big star the Sun.
The velocity field   of  the expansion modified 
by the shear velocity  is mapped.
The Galaxy's direction of  rotation is also  shown.  
}
\label{rotation_velocity}%
\end{figure}

  The physical 
parameters concerning  the Gould Belt    as  deduced in 
\cite{Perrot_2003},
  are reported in  Table \ref{tab:gould}.
   \begin{table}
      \caption{Data of the super-shell associated with the Gould Belt}
         \label{tab:gould}
      \[
         \begin{array}{cc}
            \hline
            \noalign{\smallskip}
\mbox {Size (pc}^2)                    & 466 \cdot 746 \mbox{~at~b=0} \\
\mbox {Expansion~velocity~(km~s}^{-1})  & 17                \\
\mbox {Age~(10}^7\,{\mathrm{yr}})                  & 2.6              \\
\mbox {Total energy~\,(10}^{51}{\mathrm{erg}})      & 6               \\
            \noalign{\smallskip}
            \hline
         \end{array}
      \]
   \end{table}
The total energy  is  such as to produce  results  comparable
with the observations and  the  kinematic  age  is 
the same as in \cite{Perrot_2003,Zaninetti2007}.
In order to obtain 
$E_{\mathrm{tot}}=6.~10^{51}{\mathrm{erg}}$ 
with   $t^{\mathrm{burst}}=0.015~10^7~{\mathrm{yr}}$,
we have inserted 
$N^*$=~2000.
The time necessary to cross the Earth's orbit, that lies
104 pc away from the Belt center, 
turns out 
to be  $0.078~10^7~yr$, which means $2.52~10^7~yr$ from now.
The 2D  cut at {\it z}=0 of the SB  can be
visualized in~Figure \ref{rotation_gould_sun}.
%% beginning figure rotation_gould_sun
%% end figure rotation_gould_sun
Our thermal model gives a radial velocity at {\it z} =0  
of $V_{\mathrm {theo}}$=3.67\,$\mbox {km~s}^{-1}$.
The influence of the Galactic  rotation on the direction and
modulus of the field  of the  radial velocity
is obtained by an application of 
transformation~(\ref{vrotationpc}),
see  Figure \ref{rotation_velocity}.
A comparison should be made with figure 5 
and figure 9  in \cite{Perrot_2003}.

%% beginning figure rotation_velocity
%% end figure rotation_velocity

\subsection{Cold GW~46.4+5.5}

\begin{figure}
  \begin{center}
\includegraphics[width=7cm]{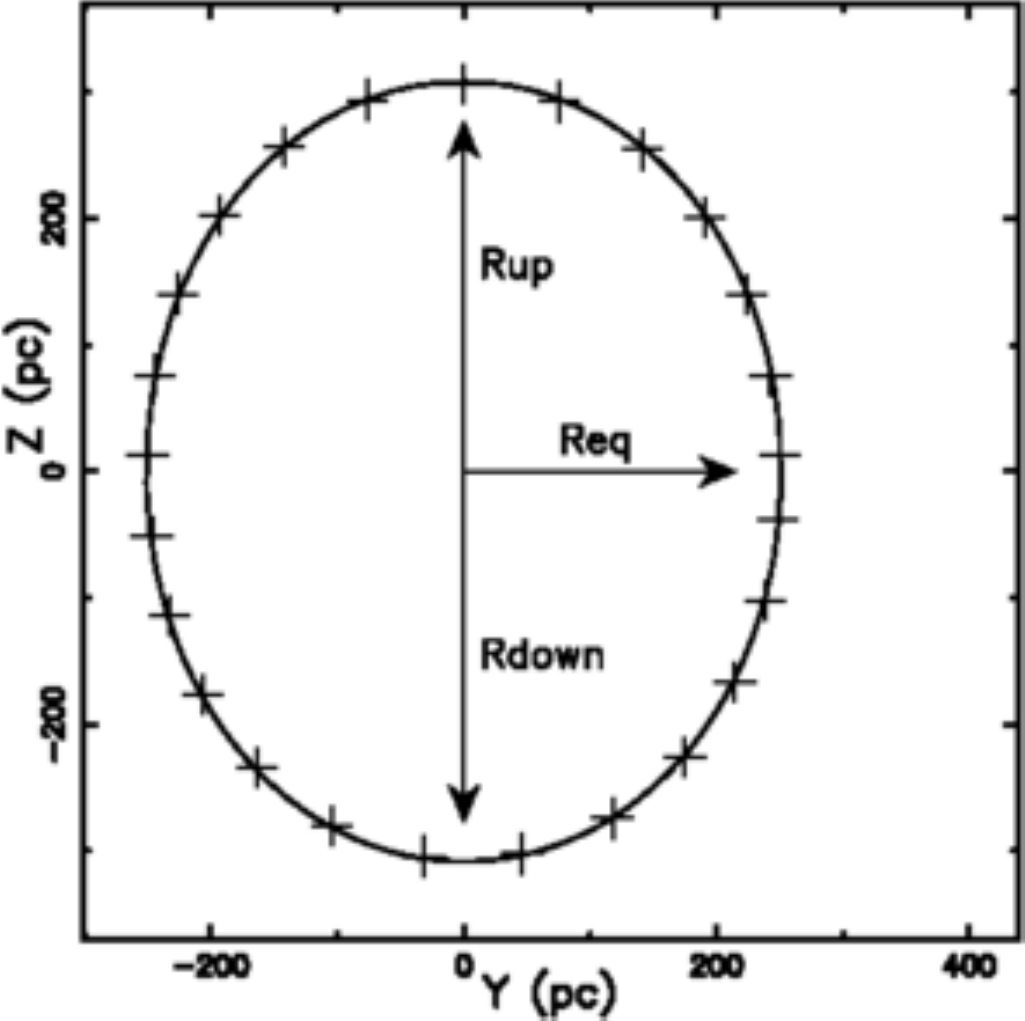}
  \end {center}
\caption
{
Section of the SB GW~46.4+5 in the {\it Y-Z;X=0}  
plane when the explosion starts at  $z_{\mathrm{OB}}=0 \mbox{pc} $.  
The analytical results  are shown as a full 
line and the numerical 
results as crosses.  
The cold code parameters of the solution of 
Eq.~(\ref{fundamental}) as well of the  numerical couple  
(\ref{recursive})
are 
$h=90$ pc, 
$t_7=0.45~$,
$t_{7,0}$ = 0.0045, 
$r_0$ = 49.12, 
$V_0=641.7\, \mathrm{km}\,\mathrm{s}^{-1}$,
$N_{SN}$ = 93   
 and  $N^*$=103000.}
\label{cut_analytical}%
    \end{figure}

\begin{figure}
  \begin{center}
\includegraphics[width=7cm]{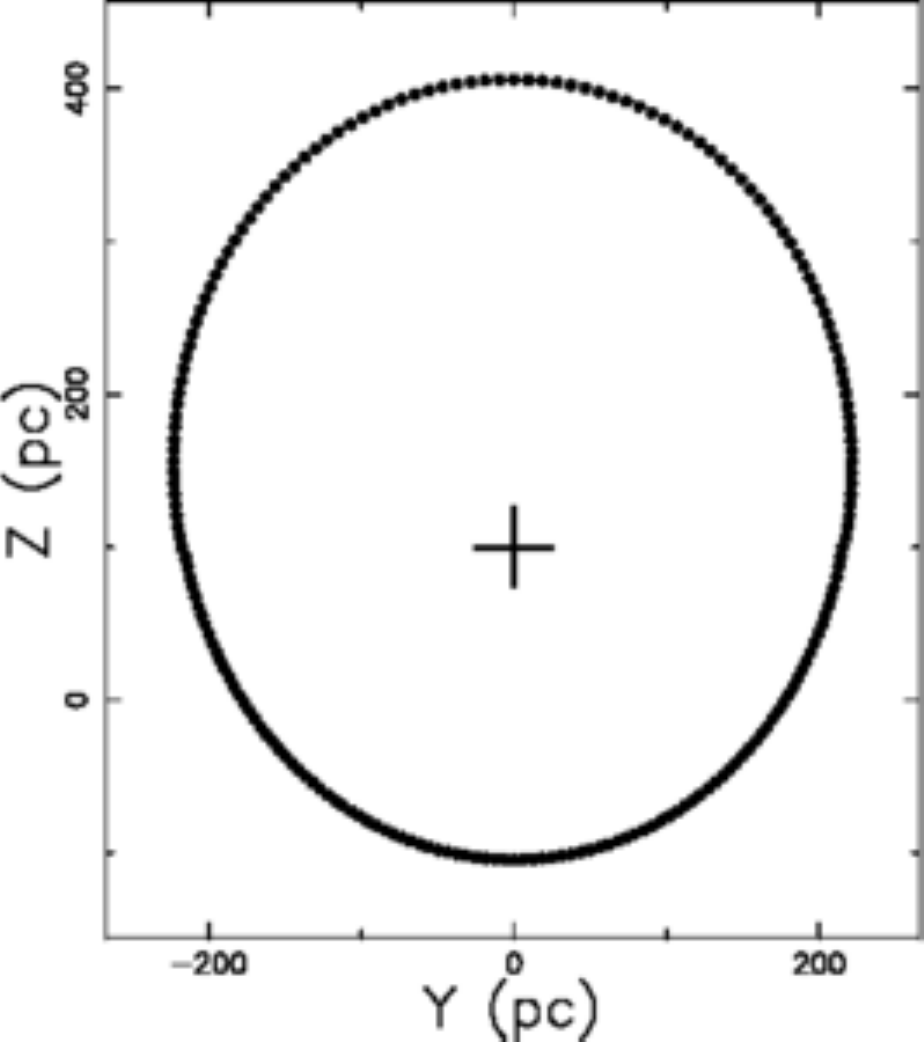}
  \end {center}
\caption
{
Section of the SB 
GW~46.4+5
in the {\it Y-Z;X=0}  plane
when the explosion starts at  $z_{\mathrm{OB}}=100 \mbox{pc}$.
The cold code parameters for  the  numerical couple  
(\ref{recursive})
are 
$h=90$ $pc$, 
$t_7=0.45~$,
$t_{7,0}$ = 0.00045, 
$r_0$ = 24.43, 
$V_0=3191 \mathrm{km}\, \mathrm{s}^{-1}$,
$N_{SN}$ = 180  
 and  $N^*$=2000000.
The explosion site is represented by a cross.
}
\label{cut_finale}%
    \end{figure}

The analytical   cold model as  
given by the solution of the  nonlinear  
Eq. (\ref{fundamental}) can be used  only  in the
case $z_{\mathrm{OB}}$=0.
As an example,  we give a model of the 
SB associated with GW~46.4+5.5, 
see  Figure  \ref{cut_analytical}.
%inizio figure cut_analytical
%fine figure cut_analytical
The  numerical  solution as given by 
the recursive relation (\ref{recursive})
can be found adopting the same  input data
as  the analytical solution, 
see the crosses  in  Figure  (\ref{cut_analytical});
the numerical solution agrees with the analytical 
solution within 0.65\%.
We  are now ready to present the numerical evolution 
of the SB   associated with GW~46.4+5
when $z_{\mathrm{OB}}=100 \mbox{pc} $,
see Figure  \ref{cut_finale} and
Table \ref{tab:ssh:simulated}.
%inizio figure cut_finale
%fine figure cut_finale
   \begin{table}
      \caption{Simulated data of the SB 
         associated with GW~46.4+5.5.}
         \label{tab:ssh:simulated}
      \[
         \begin{array}{cc}
            \hline
            \noalign{\smallskip}
\mbox {Size~(pc}^2)                    & 454 \cdot 521  \\
\mbox {Averaged expansion~velocity~ (km~s$^{-1}$}) & 19.3              \\
\mbox {Age~(10$^7$~yr})                    & 0.45             \\
\mbox {z$_{OB}$  (pc)}                   & 100             \\
            \noalign{\smallskip}
            \hline
         \end{array}
      \]
   \end{table}

\subsection{The thermal  Galactic Plane}

\label{sec_many}
\begin{figure}
\includegraphics[width=7cm]{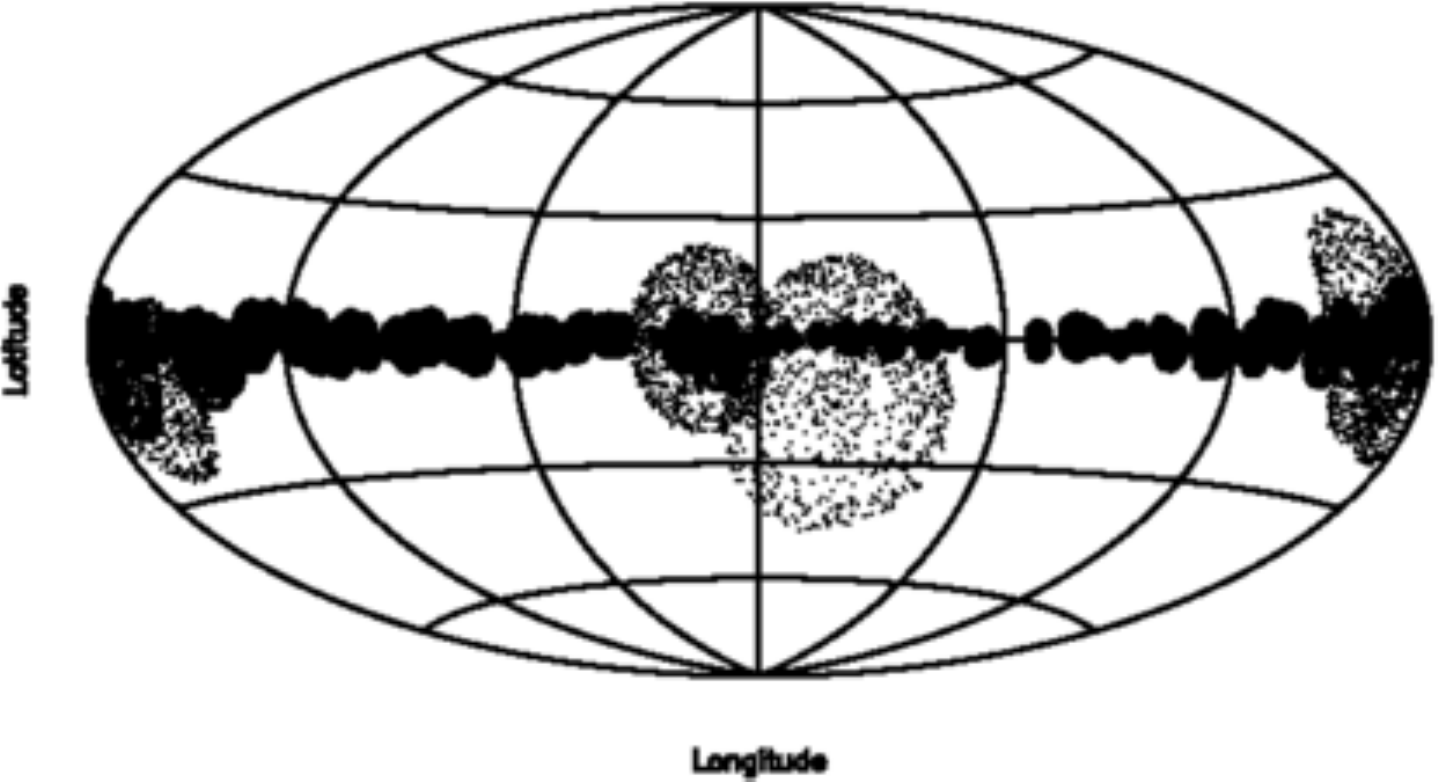}
\caption
{
Structure of  the galactic plane in the Hammer--Aitoff  projection,
as resulting from the SB/percolation  network.
The value  of  pselect  is 0.08,  corresponding to 260  selected
clusters.
The parameters   were 
$\Delta t=0.01 \cdot 10^{7}~\mbox{yr}$,
$t^{\mbox{burst}}_7$=0.5, $N^*$=  100 and each SB had
200 random points on its surface.
}
\label{plane_hamm}%
    \end{figure}

In order  to simulate the structure
of the galactic plane,
 the same  basic  parameters  as in
Figure \ref{milky} can be chosen,
but now    the SBs are drawn   on
an equal-area Aitoff projection.
 In particular,
a certain  number of  clusters
 will be selected  through  a random process
  according  to the following formula:
\begin{equation}
selected~clusters = pselect  \cdot  number~of~clusters~from ~percolation
, 
\end{equation}
where   pselect  has  a  probability  lower than one.
The  final  result  of the  simulation for the thermal case 
is reported   in Figure \ref{plane_hamm}.

\section{Theory  of the Image}

\label{sectionimage}
The transfer equation in the presence of emission
only, see for
example \cite{rybicki} or \cite{Hjellming1988}, is
 \begin{equation}
\frac {dI_{\nu}}{ds} =  -k_{\nu} \zeta I_{\nu}  + j_{\nu} \zeta
\label{equazionetrasfer} \quad,
\end {equation}
where  $I_{\nu}$ is the specific intensity, 
$s$  is the line of sight, 
$j_{\nu}$ the emission coefficient, $k_{\nu}$   a mass
absorption coefficient, $\zeta$ the  mass density
at position $s$,
and the index $\nu$ denotes the relevant
frequency of emission.
The solution to  Eq.~(\ref{equazionetrasfer})
 is
\begin{equation}
 I_{\nu} (\tau_{\nu}) =
\frac {j_{\nu}}{k_{\nu}} ( 1 - e ^{-\tau_{\nu}(s)} ) \quad ,
\label{eqn_transfer}
\end {equation}
where $\tau_{\nu}$ is the optical depth 
at frequency $\nu$
\begin{equation}
d \tau_{\nu} = k_{\nu} \zeta ds \quad.
\end {equation}
We now continue analysing the case of an
optically thin layer
in which $\tau_{\nu}$ is very small
( or $k_{\nu}$  very small )
and the density  $\zeta$ is replaced with our number density
$C(s)$ of  particles.
One case is taken into account:   the
emissivity is proportional to the number density
\begin{equation}
j_{\nu} \zeta =K  C(s) \quad ,
\end{equation}
where $K$ is a  constant.
This  can be the case for synchrotron emission.
We select  
as an example the  [S II] continuum  of 
the synchrotron SB
in the irregular galaxy
IC10, see \cite{Lozinskaya2008},
 and the 
X-ray emission below 2 keV  
around the OB association LH9 
in the H II complex N11 in the Large
Magellanic Cloud, see \cite{Maddox2009}.
The intensity at 
a given frequency 
is 
\begin{equation}
I (\nu) \propto  l  \nu^{\beta }
\quad, 
\end {equation} 
where $l$ is the length of the radiating region 
along the line of sight.
The synchrotron luminosity
is assumed  to be proportional to 
the flow  of kinetic energy
$L_m$,
\begin{equation}
L_m = \frac{1}{2}\rho A  V^3
\quad,
\label{fluxkineticenergy}
\end{equation}
where $A$ is the considered area, see formula (A28)
in \cite{deyoung}.
In our case,    
$A=4\pi R^2$,
which means
\begin{equation}
L_m = \frac{1}{2}\rho 4\pi R^2 V^3
\quad,
\label{fluxkinetic}
\end{equation}
where $R$  is the instantaneous radius of the SB  and
$\rho$  is the density in the advancing layer
in which the synchrotron emission takes place.
The astrophysical  version of the 
the rate of kinetic energy is
\begin{equation}
L_{ma} =
{ 1.39\times 10^{29}}\,{\it n_1}\,{{\it R_1}}^{2}{{\it V_1}}^{3}
\frac{ergs}{s}
\quad,
\label{kineticfluxastro}
\end{equation}
where $n_1$   is the   number density expressed
in units  of  $1~\frac{particle}{cm^3}$,
$R_1$  is  the  radius in parsecs,
and
$V_{1}$ is the   velocity in
km/s.
The spectral luminosity, $ L_{\nu} $,
at a given frequency $\nu$
is
\begin{equation}
L_{\nu} =  4 \pi  D^2  S_{\nu}
\quad ,
\end{equation}
with
\begin{equation}
S_{\nu} =  S_
0  (\frac{\nu}{\nu_0})^{\beta}
\quad ,
\end{equation}
where  $S_0$   is the flux
observed at  the frequency
$\nu_0$  and  $D$ is the  distance.
The total observed synchrotron 
luminosity, $L_{tot} $,
is
\begin{equation}
L_{tot} =
\int_{\nu_{min}}^{\nu_{max}}  L_{\nu} d \nu
\quad ,
\end{equation}
where
${\nu_{min}}$ and
${\nu_{max}}$
are the  minimum and maximum frequencies  observed.
The  total observed
luminosity
can  be expressed as
\begin{equation}
L_{tot} = \epsilon  L_{ma}
\label{luminosity}
\quad ,
\end{equation}
where  $\epsilon$  is  a constant  of conversion
from  the mechanical luminosity   to  the
total observed luminosity in synchrotron emission.

\begin{figure*}
\begin{center}
\includegraphics[width=7cm]{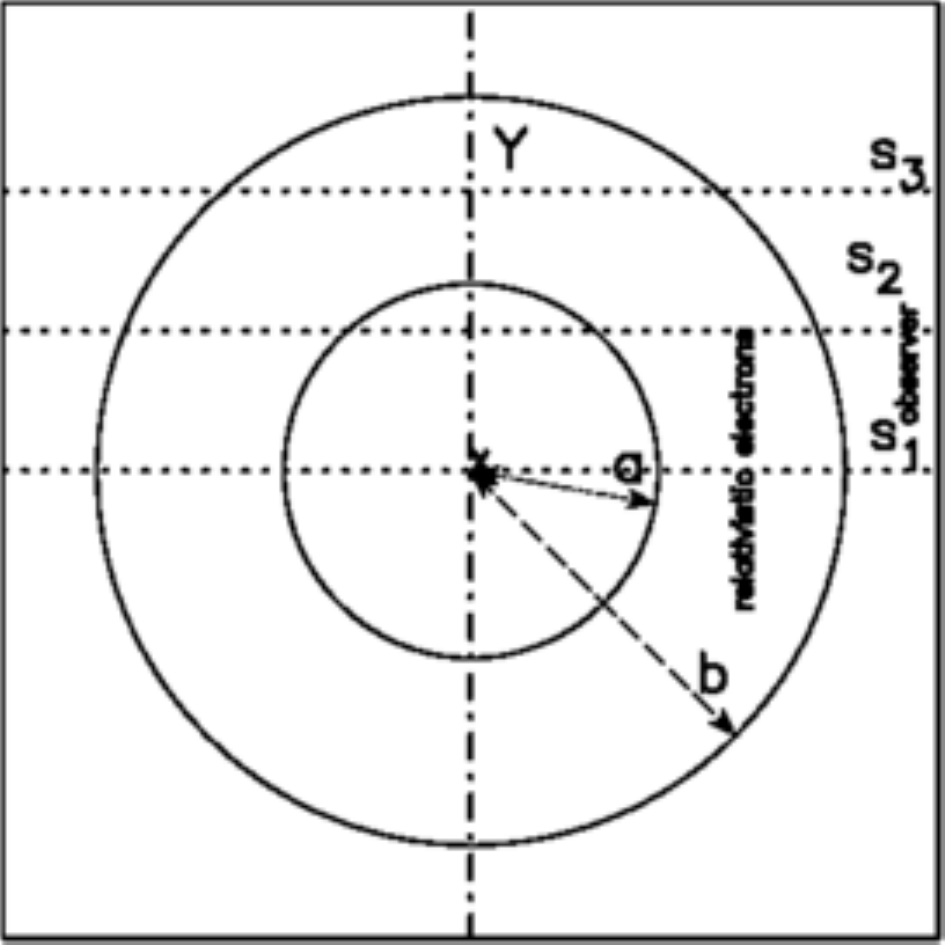}
\end {center}
\caption
{
The two circles (sections of spheres)  which   
include the region
with constant density
are   represented by
 a full line.
The observer is situated along the $x$ direction,  
three lines of sight are indicated,
and  the relativistic  electrons have  radius $r$  
in the region   $a <r <b$.
}
\label{disegnoab}%label
    \end{figure*}

The fraction  of the total  luminosity deposited  in a
band  $f_c$  is
\begin{equation}
f_c  =
\frac
{
{{\it \nu_{c,min}}}^{\beta+1}-{{\it \nu_{c,max}}}^{\beta+1}
}
{
{{\it \nu_{min}}}^{\beta+1}-{{\it \nu_{max}}}^{\beta+1}
}
\quad ,
\end{equation}
where  $\nu_{c,min}$  and  $\nu_{c,max}$
are the minimum and maximum frequencies  of the band.
Table \ref{tablecolors} shows some values of 
$f_c$  for the most important optical bands.
%tablecolors
%tablecolors
\begin{table} [h!]
\label{tablecolors}
\begin{center}
      \caption
      {
      Table of the values of $f_c$ when
      $\nu_{min}= 10^7 Hz$,
      $\nu_{max}= 10^{18} Hz$
      and  $\beta=-0.7$.
         }
         \begin{tabular}{crrr}
            \hline
           \hline
band         & $\lambda$  ($\AA$) &  FWHM  ($\AA$) & $f_c$\\
            \hline
U            &  3650   &  700  & 6.86 $\times  10^{-3}$  \\
B            &  4400   &  1000 & 7.70 $\times  10^{-3}$   \\
V            &  5500   &  900  & 5.17 $\times  10^{-3}$  \\
$ H\alpha$   &  6563   &  100  & 0.56 $\times  10^{-3}$  \\
$[S II]$ continuum   &  7040   &  210  & 0.92 $\times  10^{-3}$  \\
            \hline
            \hline
         \end{tabular}
   \end{center}
   \end{table}
%tablecolors
%tablecolors
%finemodifica
An analytical  solution  for the radial cut of intensity 
of  emission can be found  in the equatorial plane 
$z_{\mathrm{OB}}$=0.
We assume that the number density 
of relativistic electrons 
$C$ is constant and in particular
rises from 0 at $r=a$ to a maximum value $C_m$, remains
constant up  to $r=b$, and then falls again to 0,
see \cite{Zaninetti2009a}.
This geometrical  description is shown in  
Figure  \ref{disegnoab}.
% figure  disegnoab
% end figure  disegnoab
The length of sight when the observer is situated
at the infinity of the $x$-axis 
is the locus    
parallel to the $x$-axis which  crosses  the position $y$ 
in a 
Cartesian $x-y$ plane and 
terminates at the external circle
of radius $b$.
When the number density
of the relativistic electrons 
$C_m$ is constant between the two spheres
of radii $a$ and $b$, 
the intensity of the radiation is 
\begin{eqnarray}
I_{0a} =C_m \times 2 \times ( \sqrt { b^2 -y^2} - \sqrt {a^2 -y^2}) 
\quad  ;   0 \leq y < a  \nonumber  \\
I_{ab} =C_m \times  2 \times ( \sqrt { b^2 -y^2})  
 \quad  ;  a \leq y < b    \quad . 
\label{irim}
\end{eqnarray}
The ratio between the theoretical intensity 
 at the maximum      ($(y=a)$)
 and at the minimum  ($y=0$)
is given by 
\begin{equation}
\frac {I(y=a)} {I(y=0)} = \frac {\sqrt {b^2 -a^2}} {b-a}
\quad .
\label{ratioteorrim}
\end{equation}

\begin{figure*}
\begin{center}
\includegraphics[width=7cm]{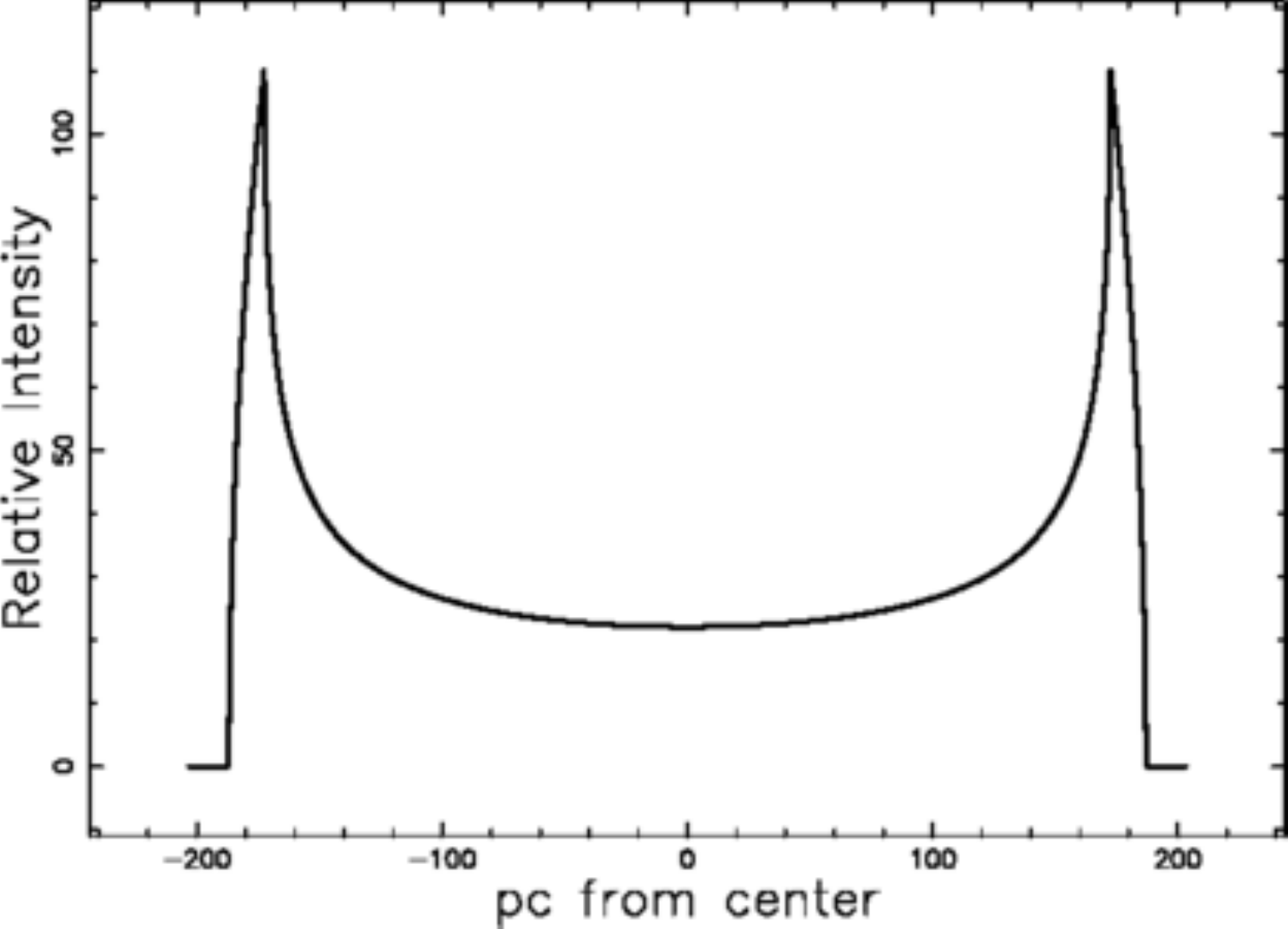}
\end {center}
\caption
{
 Cut of the mathematical  intensity ${\it I}$
 of the ring  model, Eq.~(\ref{irim}), 
 in the equatorial plane    
 (full  line) 
of the SB associated with GW~46.4+5.5~.
The $x$- and  $y$-axes  are in pc,
 $a=172.5 $ pc,  $b=186.9 $ pc
 and $\frac {I(y=a)} {I(y=0)}=5$.
}
\label{ring_cut}
    \end{figure*}

\begin{figure}
  %\begin{center}
\includegraphics[width=7cm]{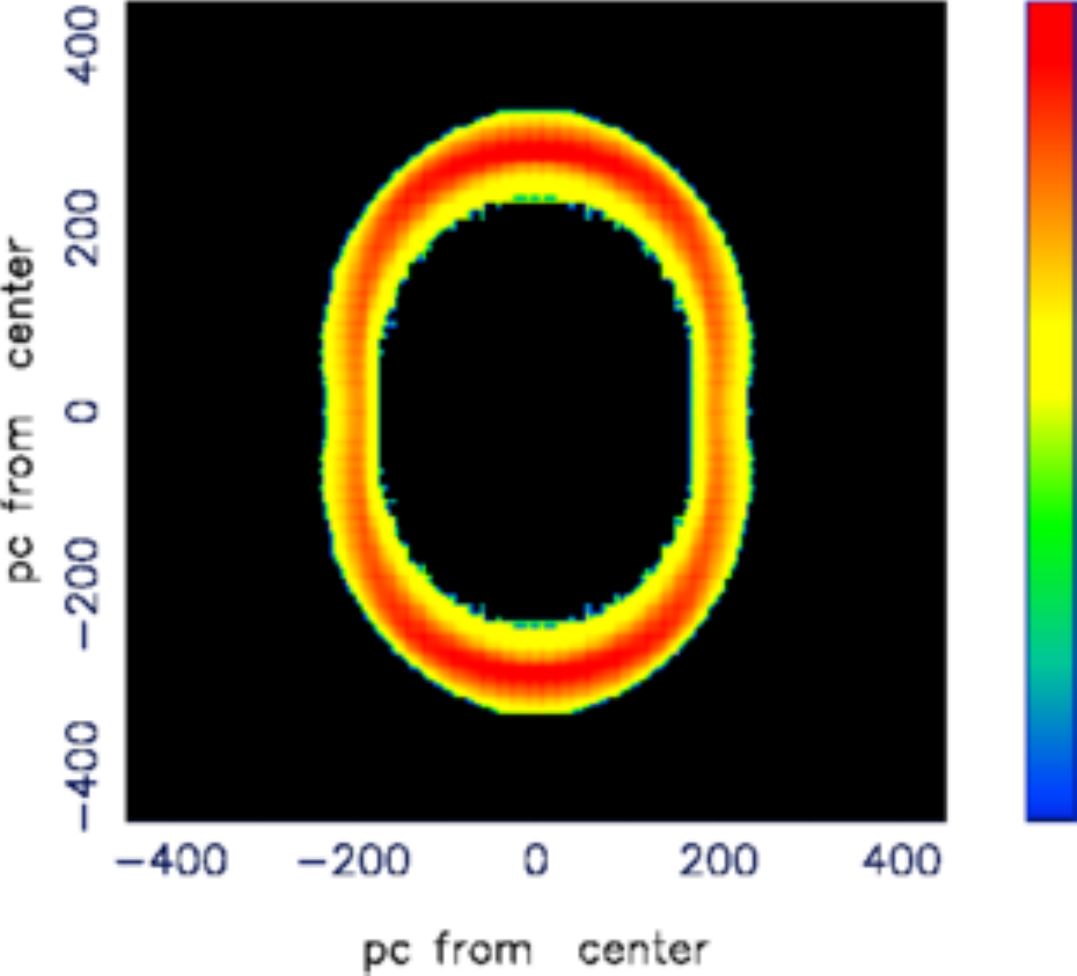}
  %\end{center}
\caption {
Map of the theoretical intensity  of
SB GW~46.4+5.
Physical parameters as in Figure  \ref{cut_analytical}.
The three Euler angles
characterizing the   orientation
  are $ \Phi  $=90 $^{\circ }$,
      $ \Theta$=90 $^{\circ }$
and   $ \Psi  $=90 $^{\circ }$.
This  combination of Euler angles corresponds
to the rotated image with the polar axis along the
$z$-axis.
In this map $I_{tr}= I_{max}/2$.
}%
    \label{sb_lum_hole}
    \end{figure}

\begin{figure}
  %\begin{center}
\includegraphics[width=7cm]{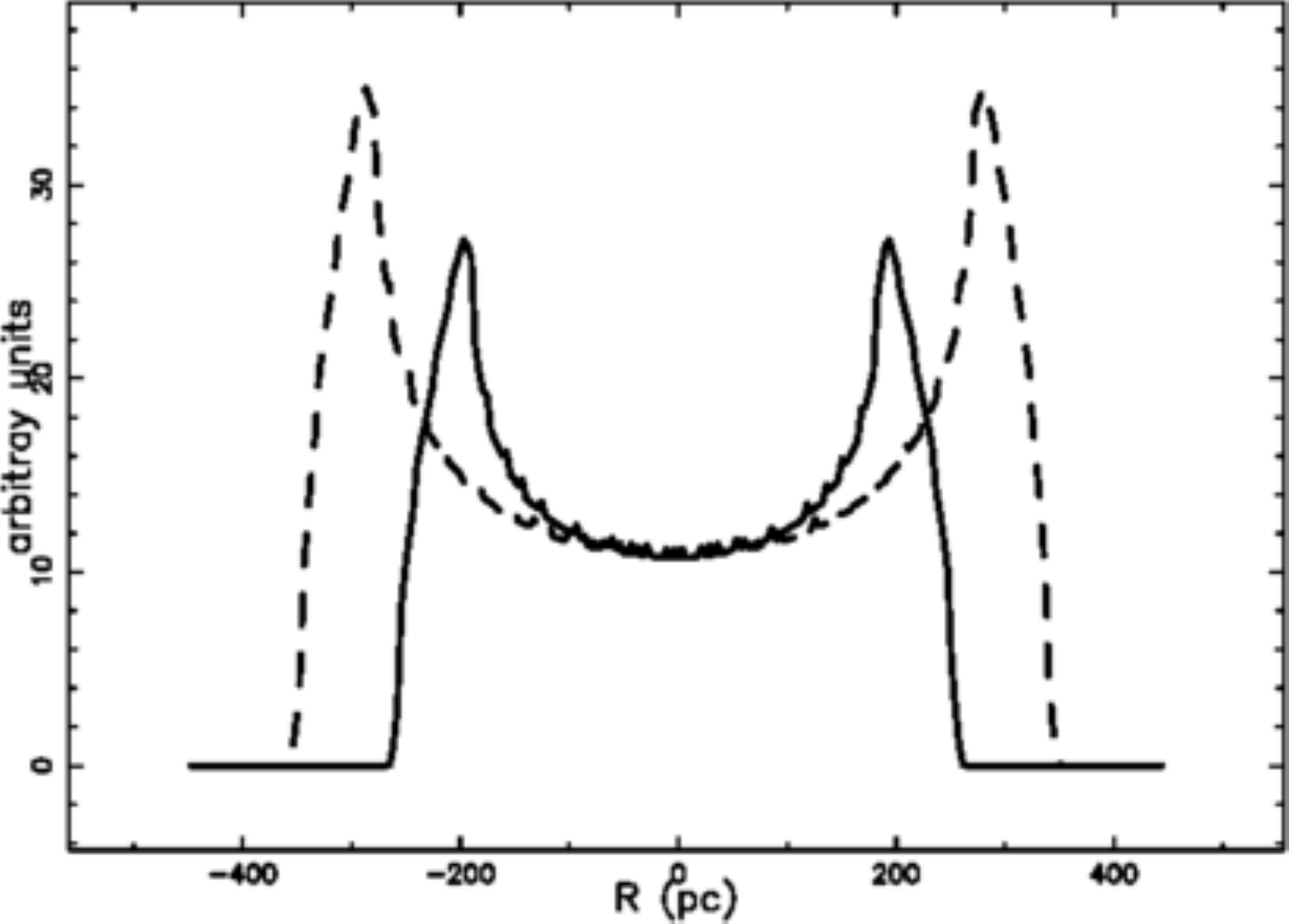}
  %\end{center}
\caption 
{
 Two cuts of the model intensity
 across the center of the SB:
 equatorial cut (full line)
 and polar cut  (dotted line).
 Parameters as in Figure   \ref{sb_lum_hole} and
 $I_{tr}= 0$.
}
    \label{cut_xy_ob}
    \end{figure}

A cut in the theoretical intensity 
of the SB associated with GW~46.4+5.5 
is shown in Figure  \ref{ring_cut}.
% figure   ring_cut
% end ring_cut
Similar analytical results for the 
intensity of the cuts 
in the H$\alpha$ region  of planetary nebulae
and in the radio emission  of supernova remnants 
have  been found
by \cite{Gray2012},
compare their figure 5 with our
Figure \ref{disegnoab}, 
 and by \cite{Opsenica2011}, see their figure 1.
A simulated image of the complex shape of an  SB 
is composed
by combining the intensities
which  characterize
different points of the advancing shell.
For an optically thin medium, the transfer equation
provides the emissivity to be multiplied by the distance
of  the line of sight, $l$.
This length, in  an SB, depends
on the  orientation of the observer,
but for the sake of clarity the observer is at infinity
and sees  the SB from the equatorial plane
$z_{\mathrm{OB}}$ =0
or from one of the two poles.  
We now outline 
the numerical algorithm which allows us to
build  the  complex  image  of an SB.
\begin{itemize}
\item 
An empty (value=0)
memory grid  ${\mathcal {M}} (i,j,k)$ which  contains
$NDIM^3$ pixels is initialized.
\item 
We  first  generate an
internal 3D surface by rotating the section of 
 $180^{\circ}$
around the polar direction and 
a second  external  surface at a
fixed distance $\Delta R$ from the first surface. 
As an example,
we fixed $\Delta R$ = $ 0.03 R_{max}$, 
where $R_{max}$ is the
maximum radius of expansion.
The points on
the memory grid which lie between the internal and the external
surfaces are memorized on
${\mathcal {M}} (i,j,k)$ with a variable integer
number   according to formula
(\ref{fluxkinetic})  and   density $\rho$ proportional
to the swept    mass.
\item Each point of
${\mathcal {M}} (i,j,k)$  has spatial coordinates $x,y,z$ 
which  can be
represented by the following $1 \times 3$  matrix, $A$,
\begin{equation}
A=
 \left[ \begin {array}{c} x \\\noalign{\medskip}y\\\noalign{\medskip}{
\it z}\end {array} \right]
\quad  .
\end{equation}
The orientation  of the object is characterized by
 the
Euler angles $(\Phi, \Theta, \Psi)$
and  therefore  by a total
$3 \times 3$  rotation matrix,
$E$, see \cite{Goldstein2002}.
The matrix point  is
represented by the following $1 \times 3$  matrix, $B$,
\begin{equation}
B = E \cdot A
\quad .
\end{equation}
\item
The intensity map is obtained by summing the points of the
rotated images
along a particular direction.
\item
The effect of the  insertion of a threshold intensity, $I_{tr}$,
given by the observational techniques,
is now analyzed.
The threshold intensity can be
parametrized by $I_{max}$,
the maximum  value  of intensity
which  characterizes  the map,
see  \cite{Zaninetti2012b}. 
\end{itemize}
An  ideal image of the intensity of  SB GW~46.4+5  
is shown in Figure  \ref{sb_lum_hole}, and 
Figure  \ref{cut_xy_ob} shows two
cuts through the center of the SB.

% figure  sb_lum_hole
% end figure sb_lum_hole

% figure  cut_xy_ob
% end figure cut_xy_ob

\begin{figure}
  %\begin{center}
\includegraphics[width=7cm]{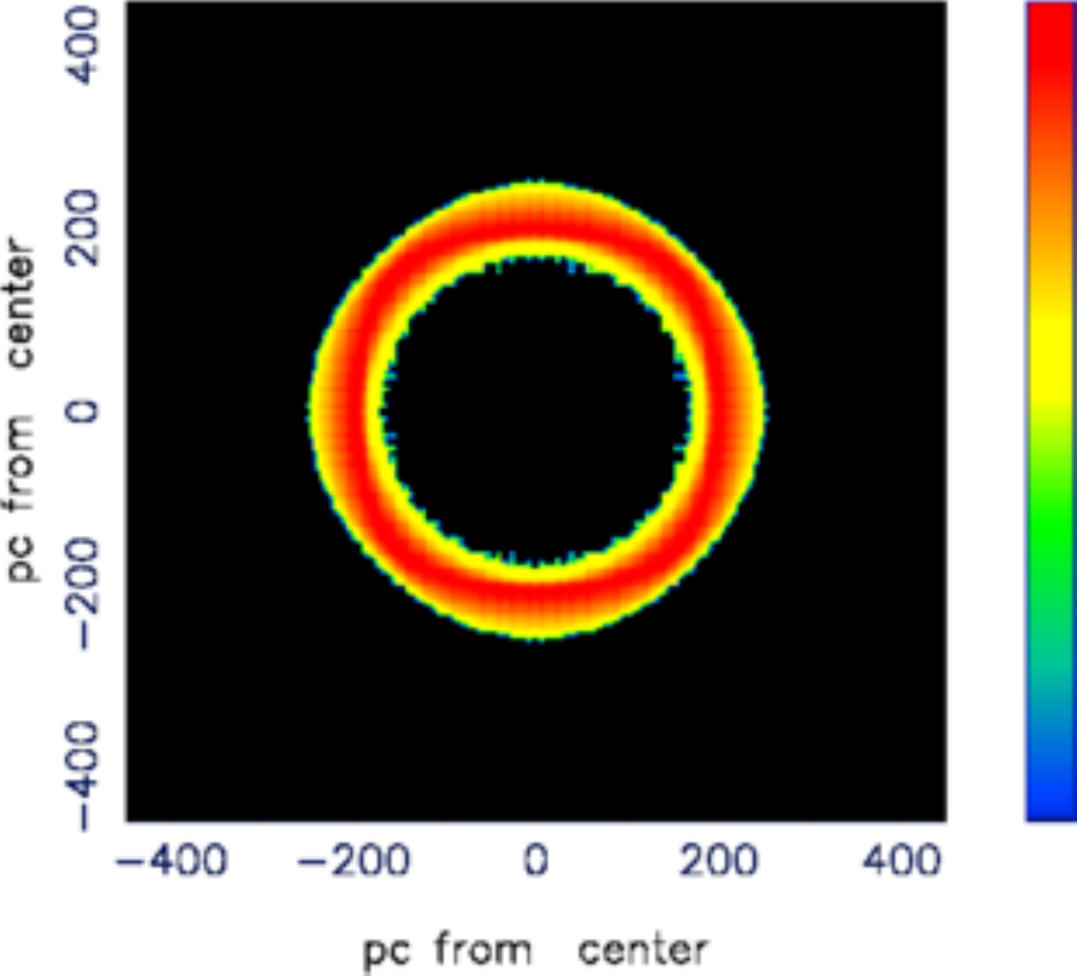}
  %\end{center}
\caption {
Map of the theoretical intensity  of
SB GW~46.4+5 as seen from the polar direction 
at infinity.
Physical parameters as in Figure  \ref{cut_analytical}.
The three Euler angles
characterizing the   orientation
  are $ \Phi  $=0$^{\circ }$,
      $ \Theta$=0$^{\circ }$
and   $ \Psi  $=0$^{\circ }$.
In this map $I_{tr}= I_{max}/2$.
}%
    \label{sb_lum_hole_polar}
    \end{figure}

We can also build the theoretical image as 
seen from one of the two
poles, see Figure  \ref{sb_lum_hole_polar}.
% figure  sb_lum_hole_polar
% end figure sb_lum_hole_polar

\begin{figure}
\includegraphics[width=7cm]{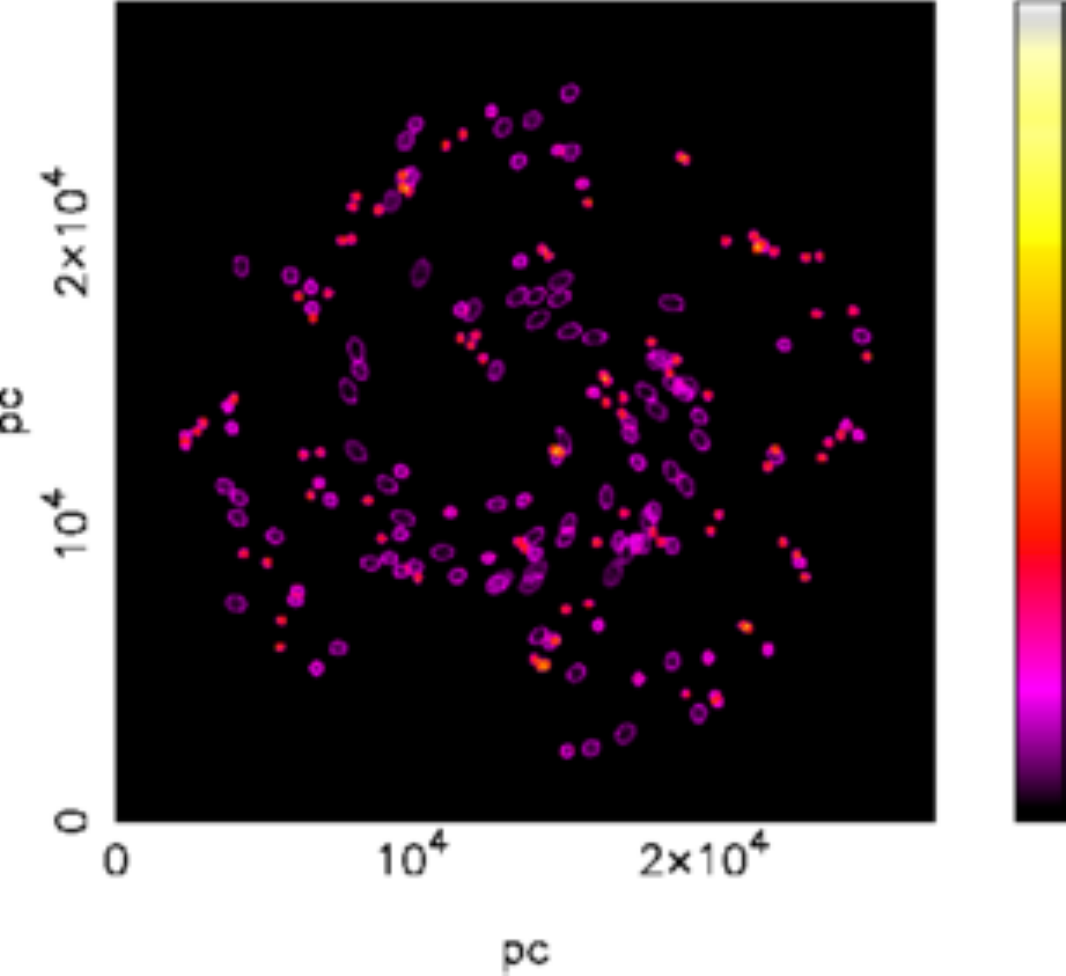}
\caption
{
 Map of the theoretical intensity  of the Milky Way 
 relative to the   network   of the explosions  when
 the galaxy is face on. 
 Physical  parameters as in
 Figure \ref{plane_hamm}.
}
\label{spiral_bubbles}%
    \end{figure}

We can also visualize the structure of the SBs
as seen from an observer situated outside the Galaxy;
the spiral structure arising from the
ensemble of the shells is evident  (see Figure \ref{spiral_bubbles}).
The elliptical  shape of the SB, 
according to 
formula~(\ref{phitotale}), is a function of both the age  and the distance
from the center. 
The distortion follows the inclination of the arms.
%figure  spiral_bubbles 
% end figure  spiral_bubbles 

\begin{figure}
\includegraphics[width=7cm]{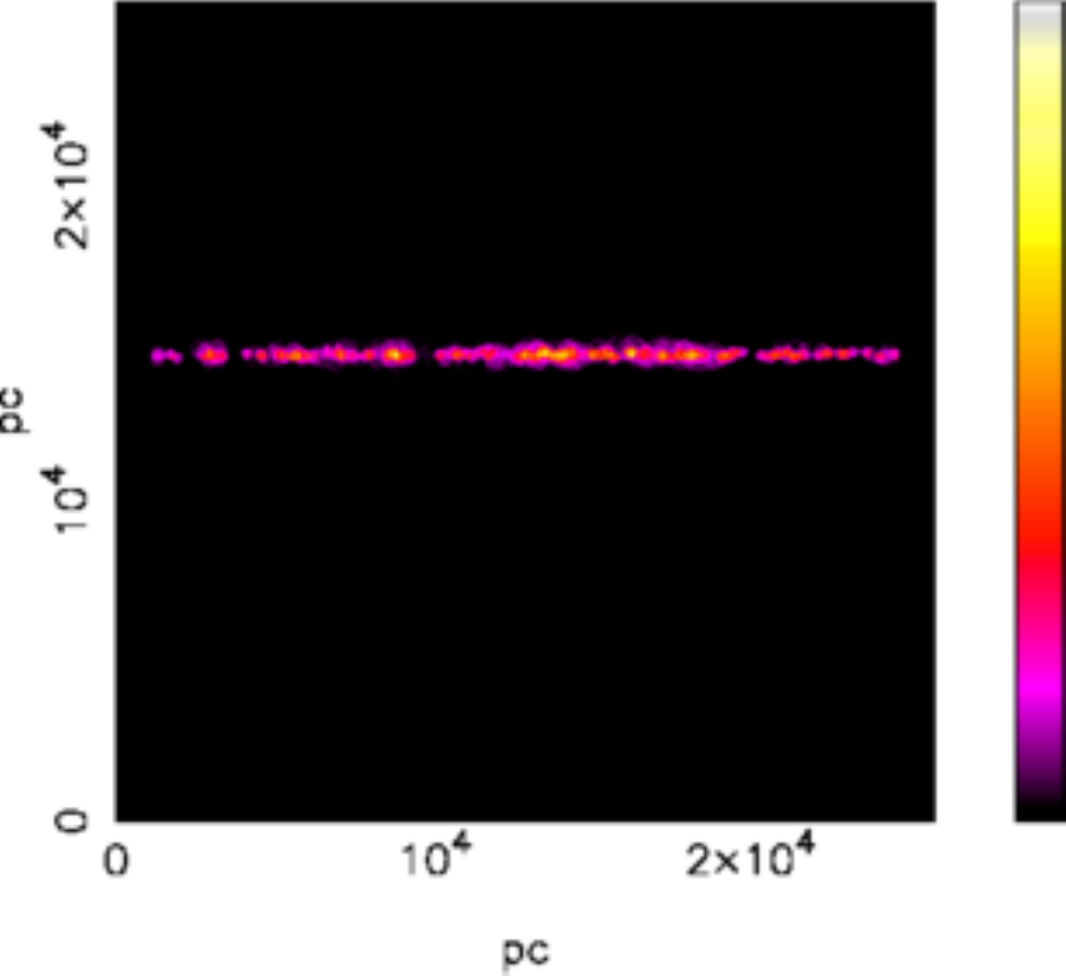}
\caption
{
 Map of the theoretical intensity  of the Milky Way 
 relative to the   network   of the explosions  when
 the galaxy is edge  on. 
 Physical  parameters as in
 Figure \ref{plane_hamm}.
}
\label{spiral_bubbles_edge}%
    \end{figure}

Figure \ref{spiral_bubbles_edge} reports 
the  image of a galaxy as seen edge on.
%figure  spiral_bubbles_edge
% end figure  spiral_bubbles_edge 

\section{Conclusions}

{\bf Percolation and Spiral Galaxies}
The active regions of the percolation theory 
for spirals are usually connected with the SNRs 
but also the SBs can trigger the formation of new stars.  

{\bf Thermal  law of motion}
The expansion of a super-bubble in the ISM belonging to our galaxy 
can be simulated by applying   Newton's second law 
to  different pyramidal  sectors. 
The following results  were  achieved:
\begin{enumerate}
\item      Single objects,  like the super-shells associated 
           with GW~46.4+5 and GSH\,238,  were simulated with an
           efficiency of $\epsilon_{\mathrm {obs}}\approx 68 \%$
            and $\epsilon_{\mathrm {obs}}\approx 58  \%$, respectively.
\item      The network of many explosions that originate from 
           the galactic plane could  be tentatively 
           simulated. 
\end{enumerate}

{\bf Cold and asymmetrical law of motion}
The temporal evolution  of  an SB  in a medium 
with constant  density is  characterized  by a
spherical symmetry.
The presence  of a thin
self-gravitating disk of gas which is characterized by a
Maxwellian distribution
in velocity and  a distribution of density which varies
only in the $z$-direction    produces 
an axial  symmetry in the temporal evolution 
of an SB.
The resulting Eq. (\ref{fundamental})  
has an 
analytical form which can be solved numerically
when $z_{\mathrm{OB}}$ =0.
The case of  $z_{\mathrm{OB}}\neq 0$ can be attached 
solving two recursive equations, see (\ref{recursive}).

{\bf Astrophysical Applications}

The thermal  model
was applied to  GW~46.4+5.5,
with an efficiency of $\epsilon_{\mathrm {obs}}=68.4\%$,
see Section \ref{secthermalgw46}, 
and to the    Gould Belt,
for which a map of the velocities as modified by
the galactic rotation is reported,
see Figure \ref{rotation_velocity}.
The cold model was applied to GW~46.4+5.5.

{\bf Images}

The combination of different processes such as
the complex shape of SBs,
the thin layer approximation
(which means an advancing layer having a thickness
$\approx 1/10$ of the momentary radius),
the conversion of the rate of kinetic energy into luminosity
and the observer's point of view allows
simulating the 
map of the theoretical intensity  of
GW~46.4+5.5, see Figure \ref{sb_lum_hole}.
The grand design of the SBs in a spiral network 
is also simulated, see Figure \ref{spiral_bubbles}.

\section*{Acknowledgments}

The credit for Figure \ref{m100} is  given to  JPL/Caltech.
At the moment of writing some animations concerning 
the spiral galaxies as  modeled by percolation 
theory are available at 
\url{http://personalpages.to.infn.it/~zaninett/animations.html}.
%\bibliography{biblio}

\begin{thebibliography}{10}
\expandafter\ifx\csname url\endcsname\relax
  \def\url#1{\texttt{#1}}\fi
\expandafter\ifx\csname urlprefix\endcsname\relax\def\urlprefix{URL }\fi

\bibitem{heiles1979}
{Heiles}, C., {H I shells and supershells}, \apj 229 (1979) 533.

\bibitem{Puche1992}
{Puche}, D., {Westpfahl}, D., {Brinks}, E., \& {Roy}, J.-R., {Holmberg II---A
  laboratory for studying the violent interstellar medium}, \aj 103 (1992)
  1841.

\bibitem{Walter1998}
{Walter}, F., {Kerp}, J., {Duric}, N., {Brinks}, E., \& {Klein}, U., {X-Ray
  Emission from an Expanding Supergiant Shell in IC 2574}, \apjl 502 (1998)
  L143.

\bibitem{PikelNer1968}
{Pikel'Ner}, S.~B., {Interaction of Stellar Wind with Diffuse Nebulae}, \aplett
  2 (1968) 97.

\bibitem{weaver}
{Weaver}, R., {McCray}, R., {Castor}, J., {Shapiro}, P., \& {Moore}, R.,
  {Interstellar bubbles. II---Structure and evolution}, \apj 218 (1977) 377.

\bibitem{Tenorio-Tagle1988}
{Tenorio-Tagle}, G. \& {Bodenheimer}, P., {Large-scale expanding
  superstructures in galaxies}, \araa 26 (1988) 145.

\bibitem{Santillian1999}
{Santill{\'a}n}, A., {Franco}, J., {Martos}, M., \& {Kim}, J., {The Collisions
  of High-Velocity Clouds with a Magnetized Gaseous Galactic Disk}, \apj 515
  (1999) 657.

\bibitem{Koo1992}
{Koo}, B.-C., {Heiles}, C., \& {Reach}, W.~T., {Galactic worms. I---Catalog of
  worm candidates}, \apj 390 (1992) 108.

\bibitem{Ambrocio-Cruz2016}
{Ambrocio-Cruz}, P., {Le Coarer}, E., {Rosado}, M., {et~al.}, {The kinematical
  properties of superbubbles and H II regions of the Large Magellanic Cloud
  derived from the 3D H{$\alpha$} Survey}, \mnras 457 (2016) 2048.

\bibitem{Sanchez-Cruces2015}
{S{\'a}nchez-Cruces}, M., {Rosado}, M., {Rodr{\'{\i}}guez-Gonz{\'a}lez}, A., \&
  {Reyes-Iturbide}, J., {Kinematics of Superbubbles and Supershells in the
  Irregular Galaxy, NGC 1569}, \apj 799 (2015) 231.

\bibitem{McCray1987}
{McCray}, R.~A., {Coronal interstellar gas and supernova remnants}, in:
  {A.~Dalgarno \& D.~Layzer} (Ed.), \textit{Spectroscopy of Astrophysical Plasmas},
  {Cambridge University Press}, Cambridge, 1987, pp.~255--278.

\bibitem{mccrayapj87}
{McCray}, R. \& {Kafatos}, M., {Supershells and propagating star formation},
  \apj 317 (1987) 190.

\bibitem{MacLow1988}
{Mac Low}, M.-M. \& {McCray}, R., {Superbubbles in disk galaxies}, \apj 324
  (1988) 776.

\bibitem{Igumenshchev1990}
{Igumenshchev}, I.~V., {Shustov}, B.~M., \& {Tutukov}, A.~V., {Dynamics of
  supershells---Blow-out}, \aap 234 (1990) 396.

\bibitem{Basu1999}
{Basu}, S., {Johnstone}, D., \& {Martin}, P.~G., {Dynamical Evolution and
  Ionization Structure of an Expanding Superbubble: Application to W4}, \apj
  516 (1999) 843.

\bibitem{Bisnovatyi-Kogan1995}
{Bisnovatyi-Kogan}, G.~S. \& {Silich}, S.~A., {Shock-wave propagation in the
  nonuniform interstellar medium}, Reviews of Modern Physics 67 (1995) 661.

\bibitem{Begelman1992}
{Begelman}, M.~C. \& {Li}, Z.-Y., {An axisymmetric magnetohydrodynamic model
  for the Crab pulsar wind bubble}, \apj 397 (1992) 187.

\bibitem{Moreno1999}
{Moreno}, E., {Alfaro}, E.~J., \& {Franco}, J., {The Kinematics of Stars
  Emerging from Expanding Shells: An Analysis of the Gould Belt}, \apj 522
  (1999) 276.

\bibitem{MacLow1989}
{Mac Low}, M.-M., {McCray}, R., \& {Norman}, M.~L., {Superbubble blowout
  dynamics}, \apj 337 (1989) 141.

\bibitem{Ferriere1991}
{Ferriere}, K.~M., {Mac Low}, M.-M., \& {Zweibel}, E.~G., {Expansion of a
  superbubble in a uniform magnetic field}, \apj 375 (1991) 239.

\bibitem{Tomisaka1992}
{Tomisaka}, K., {The evolution of a magnetized superbubble}, \pasj 44 (1992)
  177.

\bibitem{Tomisaka1998}
{Tomisaka}, K., {Superbubbles in magnetized interstellar media: Blowout or
  confinement?}, \mnras 298 (1998) 797.

\bibitem{Kamaya1998}
{Kamaya}, H., {Final Size of a Magnetized Superbubble}, \apjl 493 (1998) L95.

\bibitem{Seiden1979}
{Seiden}, P.~E. \& {Gerola}, H., {Properties of spiral galaxies from a
  stochastic star formation model}, \apj 233 (1979) 56.

\bibitem{Seiden1983}
{Seiden}, P.~E., {The role of the gas in propagating star formation}, \apj 266
  (1983) 555.

\bibitem{Schulman1986}
{Schulman}, L.~S. \& {Seiden}, P.~E., {Percolation and galaxies}, Science 233
  (1986) 425.

\bibitem{Zaninetti1988}
{Zaninetti}, L., {Percolation and synchrotron emission. I---The case of spiral
  galaxies}, \aap 190 (1988) 17.

\bibitem{Seiden1990}
{Seiden}, P.~E. \& {Schulman}, L.~S., {Percolation model of galactic
  structure}, Advances in Physics 39 (1990) 1.

\bibitem{Jungwiert1994}
{Jungwiert}, B. \& {Palous}, J., {Stochastic self-propagating star formation
  with anisotropic probability distribution}, \aap 287 (1994) 55.

\bibitem{Palous1994}
{Palous}, J., {Tenorio-Tagle}, G., \& {Franco}, J., {Star formation in
  differentially rotating galactic disks: The physics of self-propagation},
  \mnras 270.

\bibitem{Zaninetti2004}
{Zaninetti}, L., {On the Shape of Superbubbles Evolving in the Galactic Plane},
  \pasj 56 (2004) 1067.

\bibitem{press}
{Press}, W.~H., {Teukolsky}, S.~A., {Vetterling}, W.~T., \& {Flannery}, B.~P.,
  \textit{Numerical Recipes in FORTRAN. The Art of Scientific Computing}, Cambridge
  University Press, Cambridge, 1992.

\bibitem{Bisnovatyi1995}
{Bisnovatyi-Kogan}, G.~S. \& {Silich}, S.~A., {Shock-wave propagation in the
  nonuniform interstellar medium}, \rmp 67 (1995) 661.

\bibitem{Dickey1990}
{Dickey}, J.~M. \& {Lockman}, F.~J., {H I in the Galaxy}, \araa 28 (1990) 215.

\bibitem{Lockman1984}
{Lockman}, F.~J., {The H I halo in the inner galaxy}, \apj 283 (1984) 90.

\bibitem{mckee}
{McKee}, C.~F., {Astrophysical shocks in diffuse gas}, in: {Dalgarno}, A. \&
  {Layzer}, D. (Eds.), \textit{Spectroscopy of Astrophysical Plasmas}, 1987, pp.~226--254.

\bibitem{Zaninetti2012g}
Zaninetti, L., Evolution of superbubbles in a self-gravitating disc, Monthly
  Notices of the Royal Astronomical Society 425 (2012) 2343.

\bibitem{Dyson1997}
{{Dyson}, J.~E. and {Williams}, D.~A.}, \textit{The Physics of the Interstellar
  Medium}, Institute of Physics Publishing, Bristol, UK, 1997.

\bibitem{Padmanabhan_II_2001}
{Padmanabhan}, P., {\it Theoretical Astrophysics. Vol. II: Stars and Stellar
  Systems}, {Cambridge University Press}, Cambridge, {2001}.

\bibitem{Smith2002}
{Smith }, N., {Dissecting the Homunculus nebula around Eta Carinae with
  spatially resolved near-infrared spectroscopy}, \mnras 337 (2002) 1252.

\bibitem{Spitzer1942}
{Spitzer}, Jr., L., {The Dynamics of the Interstellar Medium. III. Galactic
  Distribution}, \apj 95 (1942) 329.

\bibitem{Rohlfs1977}
{Rohlfs}, K. (Ed.), {\it Lectures on Density Wave Theory}, Vol.~69 of Lecture Notes
  in Physics, Springer-Verlag, Berlin, 1977.

\bibitem{Bertin2000}
{Bertin}, G., {Dynamics of Galaxies}, {Cambridge University Press}, Cambridge,
  2000.

\bibitem{Padmanabhan_III_2002}
{Padmanabhan}, P., {\it Theoretical Astrophysics. Vol. III: Galaxies and
  Cosmology}, {Cambridge University Press}, Cambridge, 2002.

\bibitem{Hill1828}
Hill, C.~J., {Ueber die Integration logarithmisch-rationaler Differentiale},
  J. Reine Angew. Math. 3 (1828) 101.

\bibitem{Lewin1981}
Lewin, L., {\it Polylogarithms and Associated Functions}, {North-Holland}, New
  York, 1981.

\bibitem{NIST2010}
Olver, F. W. J., Lozier, D. W., Boisvert, R. F., \& Clark, C. W.,
  {NIST Handbook of Mathematical Functions}, {Cambridge University Press},
  Cambridge, 2010.

\bibitem{Kompaneets1960}
{Kompaneyets}, A.~S., {A Point Explosion in an Inhomogeneous Atmosphere},
  Soviet Phys. Dokl. 5 (1960) 46.

\bibitem{Olano2009}
{Olano}, C.~A., {The propagation of the shock wave from a strong explosion in a
  plane-parallel stratified medium: The Kompaneets approximation}, \aap 506
  (2009) 1215. 

\bibitem{Tomisaka1986}
{Tomisaka}, K. \& {Ikeuchi}, S., {Evolution of superbubble driven by sequential
  supernova explosions in a plane-stratified gas distribution}, \pasj 38 (1986)
  697.

\bibitem{Stone1992}
{Stone}, J.~M. \& {Norman}, M.~L., {ZEUS-2D: A radiation magnetohydrodynamics
  code for astrophysical flows in two space dimensions. I---The hydrodynamic
  algorithms and tests.}, \apjs 80 (1992) 753.

\bibitem{Wouterloot_1990}
{Wouterloot}, J.~G.~A., {Brand}, J., {Burton}, W.~B., \& {Kwee}, K.~K., {IRAS
  sources beyond the solar circle. II---Distribution in the Galactic warp},
  \aap 230 (1990) 21.

\bibitem{Kim2000}
{Kim}, K.-T. \& {Koo}, B.-C., {An Infrared and Radio Study of the Galactic Worm
  GW 46.4+5.5}, \apj 529 (2000) 229.

\bibitem{Igumenshchev1992}
{Igumenshchev}, I.~V., {Tutukov}, A.~V., \& {Shustov}, B.~M., {Shapes of
  Supernova Remnants}, \sovast 36 (1992) 241.

\bibitem{koo}
{Koo}, D.~C. \& {et al.}, {Deep Pencil-Beam Redshift Surveys as Probes of Large
  Scale Structures}, in: \textit{ASP Conf. Ser. 51: Observational Cosmology}, 1993,
  112--+.

\bibitem{Hartmann1997}
{Hartmann}, D. \& {Burton}, W.~B., \textit{Atlas of Galactic Neutral Hydrogen},
  Cambridge University Press, Cambridge, 1997.

\bibitem{Perrot_2003}
{Perrot}, C.~A. \& {Grenier}, I.~A., {3D dynamical evolution of the
  interstellar gas in the Gould Belt}, \aap 404 (2003) 519.

\bibitem{Zaninetti2007}
{Zaninetti}, L., {Models of Diffusion of Galactic Cosmic Rays from
  Superbubbles}, International Journal of Modern Physics A 22 (2007) 995.

\bibitem{rybicki}
{Rybicki}, G. \& {Lightman}, A., {\it Radiative Processes in Astrophysics},
  Wiley-Interscience, New York, 1991.

\bibitem{Hjellming1988}
{{Hjellming}, R.~M.}, {\it Radio Stars in Galactic and Extragalactic Radio
  Astronomy }, Springer-Verlag, New York, 1988.

\bibitem{Lozinskaya2008}
{Lozinskaya}, T.~A., {Moiseev}, A.~V., {Podorvanyuk}, N.~Y., \& {Burenkov},
  A.~N., {Synchrotron superbubble in the galaxy IC 10: The ionized gas
  structure, kinematics, and emission spectrum}, Astronomy Letters 34 (2008)
  217.

\bibitem{Maddox2009}
{Maddox}, L.~A., {Williams}, R.~M., {Dunne}, B.~C., \& {Chu}, Y.-H.,
  {Nonthermal X-ray Emission in the N11 Superbubble in the Large Magellanic
  Cloud}, \apj 699 (2009) 911.

\bibitem{deyoung}
{De Young}, D.~S., {\it The Physics of Extragalactic Radio Sources}, University of
  Chicago Press, Chicago, 2002.

\bibitem{Zaninetti2009a}
{Zaninetti}, L., {Scaling for the intensity of radiation in spherical and
  aspherical planetary nebulae}, \mnras 395 (2009) 667.

\bibitem{Gray2012}
{Gray}, M.~D., {Matsuura}, M., \& {Zijlstra}, A.~A., {Radiation transfer in the
  cavity and shell of a planetary nebula}, \mnras 422 (2012) 955.

\bibitem{Opsenica2011}
{Opsenica}, S. \& {Arbutina}, B., {Numerical Code for Fitting Radial Emission
  Profile of a Shell Supernova Remnant. Application}, Serbian Astronomical
  Journal 183 (2011) 75.

\bibitem{Goldstein2002}
{Goldstein}, H., {Poole}, C., \& {Safko}, J., {\it Classical Mechanics},
  Addison-Wesley, San Francisco, 2002.

\bibitem{Zaninetti2012b}
{Zaninetti}, L., {On the spherical-axial transition in supernova remnants},
  Astrophysics and Space Science 337 (2012) 581.

\end{thebibliography}

\end{document}